%% file: paper.tex
\documentclass[a4paper]{aa}

\usepackage{natbib}
\usepackage[english]{babel}
\usepackage{latexsym}
\usepackage{epsfig}
\usepackage{amsmath}
\usepackage{txfonts}

\input{Befehle}

\newcommand{\tX}{{t_{\rm X}}}
\newcommand{\err}{5\,}

\renewcommand{\kB}{k}
\newcommand{\sigDC}{\sigma_{\rm DC}}
\renewcommand{\Ne}{N_{\rm e}}

\newcommand{\gMoei}[1]{g_{\o{#1}}}
\newcommand{\vp}{\vek{p}}

\newcommand{\vph}{\hat{\vek{p}}}
\newcommand{\vkh}{\hat{\vek{k}}}

\newcommand{\Nc}{N_0}

\newcommand{\change}[1]{{#1}}

\begin{document}

\title{The double Compton emissivity in a mildly relativistic\\ 
  thermal plasma within the soft photon limit}

\titlerunning{The DC emissivity in a mildly relativistic thermal plasma}

\author{J. Chluba\inst{1}, S.Yu. Sazonov\inst{1,2} \and R.A. Sunyaev\inst{1,2}}
\authorrunning{Chluba, Sazonov \and Sunyaev}

\institute{Max-Planck-Institut f\"ur Astrophysik, Karl-Schwarzschild-Str. 1,
85741 Garching bei M\"unchen, Germany \and 
Space Research Institute, Russian Academy of Sciences, Profsoyuznaya 84/32, 117997 Moscow, Russia}

\offprints{J. Chluba, \\ \email{jchluba@mpa-garching.mpg.de}}
\date{Received 6 November 2006 / Accepted 20 March 2007}
\abstract
{}
{We provide simple and accurate analytic approximations for the low frequency
  double Compton emission coefficient that are applicable in a broad range
  of physical situations up to mildly relativistic temperatures.
  These approximations may be useful for checking in which circumstances
  the double Compton process is important.}
{We perform series expansions of the double Compton emission integral for low energies of
  the incident photon and electron and compare the derived analytic
  expressions with the results obtained by numerical integrations of the full
  double Compton cross section.}
{We explicitly derived analytic approximations to the low frequency double
  Compton emission coefficient for initial monochromatic photons and Wien
  spectra. We show that combining the analytic approximations given in this
  paper, an accuracy of better than \err\% over a very broad range of
  temperatures and under various physical conditions can be achieved.
  The double Compton emissivity strongly depends on the ratio of the incoming
  photon and electron energies: for hard photons and cold electrons the
  emission is strongly suppressed compared to the case of similar photon
  and electron energy, whereas in the opposite situation, i.e. hot electrons
  and soft initial photons, the emission is enhanced.
  For photons and electrons close to thermodynamic equilibrium the double
  Compton emissivity increases less rapidly with temperature than in the
  Lightman-Thorne approximation and the corrections exceed $\sim 10\%$ for
  temperatures above $4\,$keV.  }  {}


\maketitle

\section{Introduction}

Double Compton (DC) scattering is a process of the form $e + \gamma_0
\leftrightarrow e' + \gamma_1+\gamma_2$. 
\change{It} is related to Compton scattering, like Bremsstrahlung is
related to Coulomb scattering of electrons by heavy ions.
It corresponds to the lowest order correction in the fine structure constant
$\alpha$ to Compton scattering, with {\it one additional} photon in the
outgoing channel, and like thermal Bremsstrahlung it exhibits an {\it infrared
divergence} at low frequencies. In spite of these similarities, in the case of
Bremsstrahlung the cross section directly depends on the velocity of the
electron relative to the charged particle: there is {\it no} Bremsstrahlung
emission for resting electrons (and ions). On the other hand, for DC
scattering the dependence on the velocity of the electron enters only
{\it indirectly} and can be obtained as a result of special relativistic coordinate
transformations. Therefore DC emission occurs {\it whenever} there are photons
and free electrons, no matter what the temperature of the electron gas is.

It was realized by \cite{Light81}, \citet{Thorne81} and later by
\cite{pozdnyakov1983} and \cite{Svensson84} that the DC process in comparison
to Bremsstrahlung can become the main source of {\it soft} photons in
astrophysical plasmas with low baryon density and in which magnetic fields are
negligible. It has been shown that for given number densities of protons,
$N_{\rm p}$, and photons, $N_{\rm \gamma}$, the DC process is expected to
dominate over Bremsstrahlung when\footnote{ For this estimate it was assumed
  that the incident photons are distributed according to a Wien spectrum with
  \change{effective temperature $\Te\equiv\Tg$ and that the plasma is neutral,
    $N_{\rm e}\equiv N_{\rm p}$}.}
$N_{\rm p}\lesssim 10 N_{\rm \gamma}(\kB \Te/ \me c^2)^{5/2}$
is fulfilled.
\change{This implies that due to the large entropy of the Universe (there are
$\sim\pot{1.6}{9}$ photons per baryon), at early stages (redshifts
$z\gtrsim \pot{5}{5}$, i.e. at temperatures higher than
  $\sim10^6\,$K), the DC process 
%
%
becomes very important for the thermalization of possible} spectral
distortions of the cosmic microwave background (CMB) and the evolution of
chemical potential distortions after any significant release of energy in the
early Universe \citep{Suny70, Illarionov75, Illarionov75b, Danese82,
  Burigana91, Hu93}.
Other possible environments in which the DC process could be of some
importance might be found inside the sources of $\gamma$-ray bursts
or two temperature accretion disks in the vicinity of black holes, both in
close binary systems and active galactic nuclei.
Given the relevance of DC emission as a source of soft photons within the
context of the thermalization of CMB spectral distortions and hard
X-ray sources, it is important to investigate the validity of the
approximations usually applied to describe the DC process.

\change{Expressions describing} DC emission in an {\it isotropic, cold} plasma
and for low energy initial photons, i.e. small $\omega_0=h \nu_0/\me c^2$,
where $\nu_0$ is the frequency of the initial photon $\gamma_0$, \change{were}
first derived by \citet{Light81} and independently by \citet{Thorne81} within
the {\it soft photon limit}. In this approximation it is assumed that one of
the outgoing photons ($\gamma_1$ or $\gamma_2$) is {\it soft} compared to
the other.
Under these assumptions the DC \change{cross section} increases $\propto
\omega_0^2$ with photon energy \citep[see][Eq.~11-45]{Jau76}
%
\change{and the DC emissivity for broad initial photon distributions is
  proportional to the integral $I\sim\int\nu_0^4 n(\nu_0)[1+n(\nu_0)]\id
  \nu_0$, where here $n(\nu)$ is the photon occupation number \citep{Light81}.
  We give more details about the Lightman-Thorne approximation in
  Sect.~\ref{sec:ce_np}.}
%

Again presuming cold electrons and low energy incident photons but making {\it
a priori} no assumption about the emitted photon energy, \citet{Gould84}
obtained an analytic correction factor for the DC emission coefficient of
monoenergetic initial photons, which extends the Lightman-Thorne approximation
beyond the soft photon limit.
Here the corrections are directly related to the increase of the emitted
photon energy (up to energies $\sim \nu_0 /2$), while the low frequency
emission spectrum remains unchanged.

However, when the temperature of the electrons increases or when the energy of
the incident photons grows, one expects corrections to become important, an
aspect that has not been included in either of the aforementioned approaches.
The corrections due to motion of the electron can be obtained starting
with a resting electron in an anisotropic radiation field and then performing
the corresponding transformations into the frame where the electron is moving.
In contrast to this, the dependence on the energy of the incident photon is
related to the exact formulae for the DC cross section as computed using the
general theory of quantum-electrodynamics.
Here we treat the electron-photon interaction due to DC scattering
quantum-electrodynamically for moving electrons and perform all the
computations directly in the lab frame, where the photon and electron
distributions are assumed to be isotropic.

Temperature corrections to the low frequency DC emissivity for an incoming
Wien spectrum, again within the soft photon limit, were discussed by
\citet{Svensson84}. It was shown that in the mildly relativistic case the low
frequency DC emissivity increases significantly slower with temperature than
in the Lightman-Thorne approximation. However, \citet{Svensson84} only treated
one specific case and a more general extension of the Lightman-Thorne
approximation is still missing.

In this paper we wish to extend the Lightman-Thorne approach to cases
$h\nu_0\lesssim \me c^2$ and $\kB\Te\lesssim \me c^2$ for isotropic initial
photon distributions, but still within the {\it soft photon limit}.
\change{In addition, we will focus on situations when DC {\it absorption} and
  {\it stimulated emission} are negligible. The extension to cases when
  $\omega_2\sim \omega_0$ and a discussion of the full kinetic equation for
  the DC process will be left for a future work.}
We study the DC emission integral both numerically and analytically and derive
approximations to the DC emission coefficient, which are valid in a very broad
range of physical situations up to mildly relativistic temperatures.
We show that the DC emission coefficient can be expressed in terms of the
Lightman-Thorne emission coefficient times a corrections factor, which can be
regarded as very similar to the Bremsstrahlung Gaunt-factor, although it has
a completely different physical nature.
As an example, in Fig.~\ref{fig:Rashid_mono} we summarize the results for this
low frequency DC correction factor in the case of monochromatic initial
photons.  One can see that the Lightman-Thorne approximation is accurate in a
very limited range of photon energies and electron temperatures.
We also investigate the range of applicability of our approximations for
initial Wien spectra (Sect.~\ref{sec:app}). 

In the following we use $h=c=\kB=1$.

\section{DC emission for monochromatic photons and thermal
  electrons in the soft photon limit}
\label{sec:Mono}
\change{In this Section we give a detailed derivation of the approximations
describing the DC emission for monochromatic initial photons and thermal
electrons within the soft photon limit. In particular, we provide some
additional comments on the Lightman-Thorne approximation
(Sect.~\ref{sec:ce_np}) and introduce a DC correction factor
(Sect.~\ref{sec:DCfac_mon}) relative to the Lightman-Thorne formula which will
be used in the following.
See Table \ref{tab:G_mono} for overview.
}
\subsection{General formulation}
DC emission is a result of the interaction between an electron and a photon
with {\it one} additional photon in the outgoing channel:
\beq
e(P) + \gamma(K_0) \longrightarrow e(P') + \gamma(K_1) + \gamma(K_2)\nonumber
\Abst{.}
\eeq
Here $P=(E,\vek{p}),\,P'=(E',\vek{p}')$ and $K_i=(\nu_i,\vek{k}_i)$ denote the
corresponding particle four-momenta\footnote{Henceforth bold font denotes
  3-vectors.}.
The full DC scattering squared matrix element and differential cross sections
for various limiting cases were first derived by \citet{Mandl52} and may also
be found in \citet{Jau76}.

In this Section we focus our analysis only on the emission process for {\it
isotropic, monochromatic} ·\change{initial} photons, with phase space density
$n(\nu)=\Nc\,\delta(\nu-\nu_0)/\nu_0^2$,
and {\it isotropic, thermal} electrons of temperature $\Te$. Above $N_0$ in
physical units is defined by $N_{\gamma, 0}=8\pi\, N_0/c^3$, where $N_{\gamma,
0}$ is the photon number density.
In this case the change of the photon phase space density, $n_2=n(\nu_2)$, at
frequency $\nu_2$ due to DC emission can be written as
\beal
\label{eq:n2em}
\left.\pAb{n_2}{t}\right|_{\rm em}^{\rm m} = \frac{\Nc}{\nu_2^2}\int \id^3 p \int \id\Omega_0\int\id\Omega_1
\,\gMoei{0}\frac{\id \sigDC}{\id\Lambda}\,f
\Abst{.}
\end{align}
Here the DC differential cross section \citep[cf.][Eq. 11-38]{Jau76} is given
by
\beal
\label{eq:dsig_DC}
\frac{\id \sigDC}{\id \Lambda}
=\frac{\alpha\,r_0^2}{(4\pi)^2}\,\frac{\nu_1\,\nu_2}{\gMoei{0}\gamma\,\nu_0}\,
\frac{X}{(P+K_0-K_2)\cdot \hat{K}_1}
\Abst{,}
\end{align}
where $X$ may be found in the Appendix~\ref{appendix:M2} and we defined the
{\it M{\o}ller relative speed} of the incident electron and photon as
$\gMoei{i}=\hat{P}\cdot\hat{K}_i=1-\beta\,\mu_{\rm e\it i}$, with the
dimensionless electron velocity $\beta=|\vek{v}|/c=|\vph|=|\vp|/E$ and the
directional cosine\footnote{Note that in the following an additional hat above
  3- and 4- vectors indicates that they are normalized to the time-like
  coordinate of the corresponding 4-vector.} $\mu_{\rm e\it
  i}=\vph\cdot\vkh_i$.  Furthermore $\alpha=e^2/4\pi\approx 1/137$ is the fine
structure constant, $r_0=\alpha/\me\approx \pot{2.82}{-13}\,\text{cm}$ is the
classical electron radius and
%
$\gamma=\sqrt{1-\beta^2}$
denotes the Lorentz-factor of the initial electron.

In Eq. \eqref{eq:n2em} we have neglected induced effects, i.e. we assumed
$\Nc\ll \nu_0^2$ and that the electrons are non-degenerate.
The latter simplification, for sufficiently low temperatures and electron
densities (such as in the Universe for $\Te\lesssim\me$), is justified, but as
we will discuss in Sect.~\ref{sec:induced}, for broad initial photon
distributions, stimulated DC emission can become important.  However, a full
treatment of this problem is much more complicated and is beyond the scope
of this paper.
The electron phase space density may be described by a relativistic
Maxwell-Boltzmann distribution
\beq
\label{eq:relMBD}
f(E)=\frac{\Ne}{4\pi\,\me^3\,K_2(1/\theta_{\rm e})\,\theta_{\rm e}}\,e^{-E/\me\theta_{\rm e}}
\Abst{,}
\eeq
where $K_2(1/\theta_{\rm e})$ is the modified Bessel function of the second
kind \citep{Abramovitz1965}, with $\theta_{\rm e}=\Te/\me$, and where $\Ne$ is
the electron number density, such that $\Ne=\int f(E)\id^3p$. In the low
temperature limit ($\theta_{\rm e}\ll 1$) the relativistic Maxwell Boltzmann
distribution \eqref{eq:relMBD} can be handled by the expression given in
Appendix \ref{app:MB}, which is useful for both analytic and numerical
purposes.

In general the DC emission integral~\eqref{eq:n2em} has to be performed
numerically. Some comments on the numerical approach we use to solve the
multidimensional Boltzmann integrals is given in Appendix \ref{app:MC}. 
However, in the limit of low temperatures and energies of the initial photon
it is possible to derive various useful analytic approximations, which we
discuss in the following.
In practice we obtained the numerical results for the soft photon limit by
setting the frequency of the emitted photon to a very small value as compared
to the incident photon energy ($\nu_2\sim \nu_0 \times 10^{-4}$).

\subsection{Lightman-Thorne approximation}
\label{sec:ce_np}
The expression for DC emission from monoenergetic initial photons and cold
electrons as given by \citet{Light81} and independently by \citet{Thorne81}
can be deduced from the emission integral \eqref{eq:n2em} by performing a
series expansion of the DC differential cross section \eqref{eq:dsig_DC} in
lowest orders of $\nu_0\ll \me$ and $\nu_2\ll \me$ and setting $\beta=0$,
i.e. assuming that the electrons are initially at rest. In this limit
$f(p)=\Ne\,\delta(p)/4\pi p^2$ and the integration over $\id^3 p$ can be
carried out immediately.

The series expansion in lowest order of $\nu_2\ll \me$ is equivalent to using
the {\it soft photon approximation} for the DC differential cross section,
i.e. assuming in addition that $\nu_2\ll \nu_0$.
Due to energetic arguments the scattered photon frequency has to be close to
the initial photon frequency ($\nu_1\sim\nu_0$). Therefore the {\it infrared
divergence} due to the integration over the phase space volume for the photon
$\gamma_1$ is automatically avoided. In the following we use the notations
$\mu_{ij}=\vkh_i\cdot\vkh_j$ and $\phi_{ij}$ for the directional cosines and
azimuthal angles between the photons $i$ and $j$, respectively.
It then follows:
\beal
\label{eq:dsig_DC_Light}
\left.\frac{\id \sigDC^{\rm soft}}{\id \Lambda}\right|_{\rm L}
\approx
\frac{\alpha}{4\pi^2}\,r_0^2\,\frac{\omega_0^2}{\nu_2}
\,[1+\mu_{01}^2]\left[1-\mu_{01}-\frac{\Delta\mu^2}{2}\right]
\Abst{,}
\end{align}
with $\Delta\mu=\mu_{02}-\mu_{12}$. Furthermore we have introduced the
dimensionless photon energy $\omega_i=\nu_i/\me$. 
%

Now, after aligning the $z$-axis with the direction of the emitted photon all
the integrations can be easily performed analytically. Making use of the
identity
$
\mu_{01}=\mu_{02}\,\mu_{12}+\cos(\phi_{02}-\phi_{12})(1-\mu_{02}^2)^{1/2}(1-\mu_{12}^2)^{1/2}
$,
where $\phi_{02}$ and $\phi_{12}$ denote the azimuthal angles of the initial
and the scattered photon, respectively, this leads to the {\it
  Lightman-Thorne} approximation for the DC emission spectrum of cold
electrons and soft, monochromatic initial photons
\beal
\label{eq:n2em_Light}
\left.\pAb{n_2}{t}\right|_{\rm em,L}^{\rm m} = \frac{4\alpha}{3\pi}\,\Ne\,\Nc\,\sigT\,\frac{\omega_0^2}{\nu_2^3}
\Abst{,}
\end{align}
with the {\it Thomson} scattering cross section $\sigT=8\pi\,r_0^2/3\approx
\pot{6.65}{-25}\rm cm^2$.
%
The expression \eqref{eq:n2em_Light} is the DC scattering equivalent of the
{\it Kramers} formula for thermal Bremsstrahlung. The characteristics of
equation \eqref{eq:n2em_Light} can be summarized as follows: the DC emission
spectrum increases $\propto\omega_0^2$ and, as mentioned earlier, exhibits an
{\it infrared} divergence for $\nu_2\rightarrow 0$.
It is usually assumed that under physical conditions inside an astrophysical
plasma, emission and absorption balance each other below some frequency,
$\nu_{2,\rm min}$, and that energy conservation in addition introduces some
high frequency cutoff, $\nu_{2,\rm max}$. Therefore the photon production rate
increases logarithmically,
\beal
\label{eq:N2em_Light}
\left.\pAb{N_2}{t}\right|_{\rm em,L}^{\rm m} =
\frac{4\alpha}{3\pi}\,\Ne\,\Nc\,\sigT\,\omega_0^2\times\ln\left(\frac{\nu_{2,\rm max}}{\nu_{2,\rm min}}\right)
\Abst{,}
\end{align}
with the ratio of the upper to the lower frequency cutoff of the DC emission spectrum.

It is obvious that due to the DC process the interacting electrons and high
frequency photons can lose some part of their energy due to the emission of
soft photons. However, in the majority of applications the energy
losses connected with Compton heating and cooling are much larger.

\begin{figure}
\centering 
\includegraphics[width=0.96\columnwidth]
{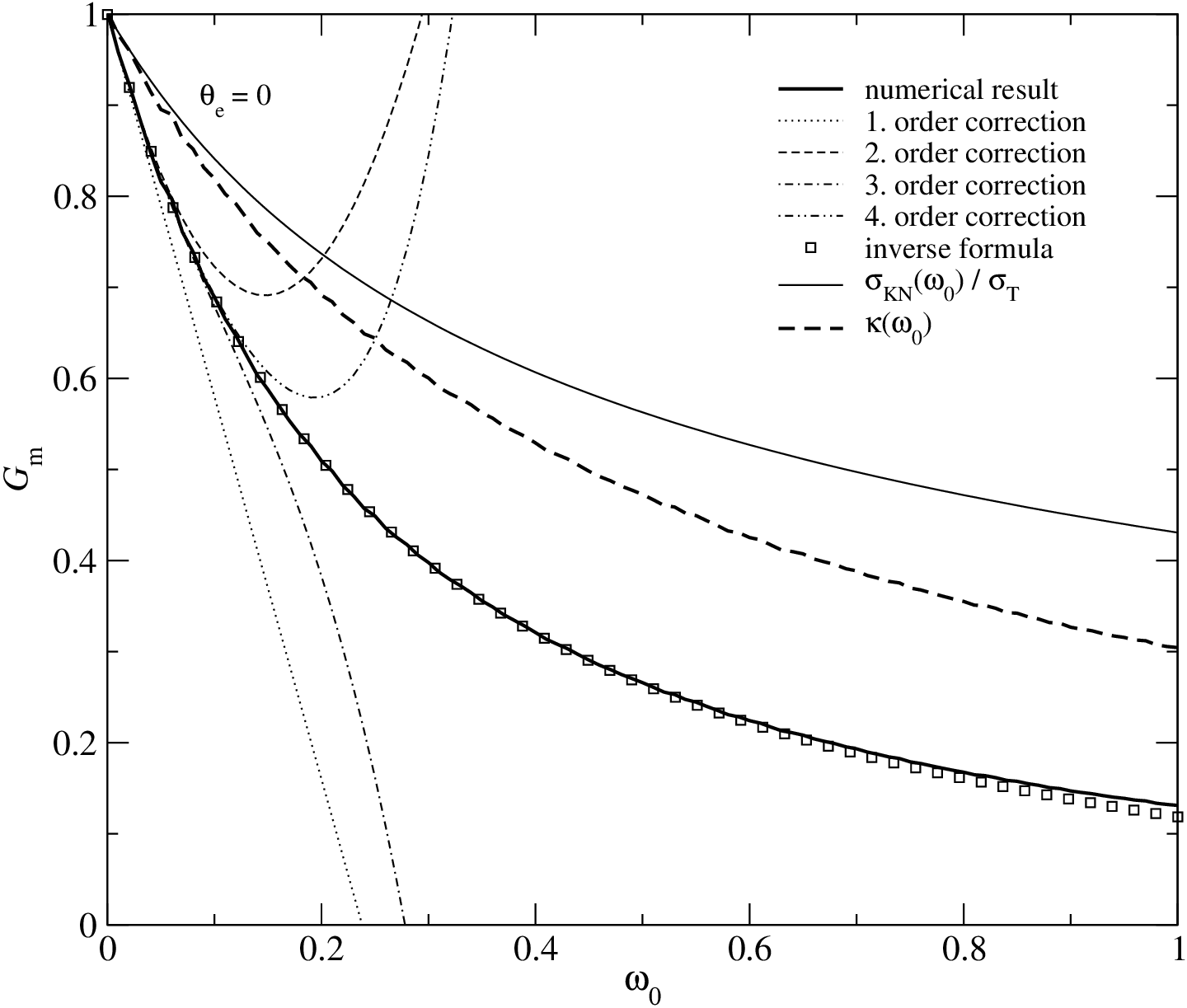}
\caption{DC emission correction factor $G^0_{\rm m}$ for cold
electrons and monochromatic initial photons as a function of
$\omega_0=\nu_0/\me$. Also shown are the {\it direct} expansion
\eqref{eq:Gmono_soft_4}, up to different orders in $\omega_0$, and the {\it
inverse} approximation \eqref{eq:G_mono_inv}.
For comparison, we also show the frequency dependence of the total Compton
cross section, $\sigma_{\rm KN}$, as given by the {\it Klein-Nishina} formula
and the correction factor $\kappa(\omega_0)=G^0_{\rm m}/(\sigma_{\rm
KN}/\sigma_{\rm T})$ per act of Compton scattering.  }
\label{fig:Gmono_soft}
\end{figure}

\begin{table}
\caption{Analytic approximations for the DC correction factor
\eqref{eq:def_Gm} relative to the Lightman-Thorne formula as obtained in
Sect.~\ref{sec:Mono}. Within the quoted range of applicability each
approximation should be accurate to better than $\sim 5\%$. However, for
$G_{\rm m}^{\rm inv}(\omega_0, \The)=\left<G^{\rm inv}_{\rm m}\right>_{\rm
th}$ as based on Eq. \eqref{eq:Gmono_soft_w0_T} and
\eqref{eq:G_mono_inv_beta_gen} we refer the reader to
Fig. \ref{fig:Rashid_mono}.  }
\label{tab:G_mono}
\centering
\begin{tabular}{@{}lccc}
\hline
\hline
Symbol & Reference & Range & comment \\
\hline
$G_{\rm m}^0(\omega_0)$ & Eq. \eqref{eq:Gmono_soft_4}, Fig. \ref{fig:Gmono_soft} 
& $\The=0$, $\omega_0\lesssim 0.15$ & expansion \\ 
$G_{\rm m}^{\rm 0, inv}(\omega_0)$ & Eq. \eqref{eq:G_mono_inv}, Fig. \ref{fig:Gmono_soft} 
& $\The=0$, $\omega_0\lesssim 1$ &  inv. factor \\
$G_{\rm m}^{\rm nr}(\The)$ & Eq. \eqref{eq:Gmono_soft_nonrel_T}, Fig. \ref{fig:Gmono_soft_nr}
& $\The\lesssim 1$, $\omega_0\ll 1$ & expansion \\
$G_{\rm m}^{\rm nr}(\The)$ & Eq. \eqref{eq:Gmono_soft_nonrel_Tb}, Fig. \ref{fig:Gmono_soft_nr} 
& $\The\lesssim 0.5$, $\omega_0\ll 1$ & expansion \\
$G_{\rm m}(\omega_0, \The)$ & Eq. \eqref{eq:Gmono_soft_w0_T_app}, Fig. \ref{fig:DG_Mono_T_w0} 
& $\The\lesssim 0.2$, $\omega_0\lesssim 0.1$ & expansion \\
$G_{\rm m}^{\rm inv}(\omega_0, \The)$ 
& see above, Fig. \ref{fig:DG_Mono_T_w0}
& Fig. \ref{fig:Rashid_mono} & inv. factor \\
\end{tabular}
\end{table}
As will be discussed in more detail below, the Lightman-Thorne approximation
only describes the low frequency part of the DC emission spectrum for cold
electrons and soft initial photons ($\nu_0/\me \leq 10^{-3}$) well. For
initial photons with slightly higher energies, next order terms lead to a
significant suppression of the emission at low frequencies relative to the
Lightman-Thorne approximation (see Fig.~\ref{fig:Gmono_soft} and the
discussion in Sect.~\ref{sec:ce_ap}).

\subsection{DC correction factor within the soft photon limit}
\label{sec:DCfac_mon}
Below we discuss the numerical and analytical results for the DC emission
integral \eqref{eq:n2em}. For convenience using Eq.~\eqref{eq:n2em} and the
Lightman-Thorne approximation \eqref{eq:n2em_Light} we introduce the new
function $G$ as
\beal
\label{eq:def_Gm}
G_{\rm m}(\omega_0, \The)
=\frac{\left.\partial_t n_2 \right|_{\rm em}^{\rm m}}{\left.\partial_t n_2 \right|_{\rm em, L}^{\rm m}}
\end{align}
in order to compare the different cases. This function can be regarded as very
similar to the Bremsstrahlung {\it Gaunt} factor, although it has a different
physical origin. Deviations of $G$ from unity are directly related to DC
higher order corrections in the energies of the initial photons and electrons.

\subsection{Cold electrons and energetic initial photons}
\label{sec:ce_ap}
We expand the DC differential cross section \eqref{eq:dsig_DC} in the lowest
order of $\omega_2=\nu_2/\me$ and setting $\beta=0$. Then we expand the
resulting expression up to 4th order in $\omega_0$ to take into account higher
order corrections in the energy of the initial photon. Carrying out all the
integrations in the emission integral \eqref{eq:n2em} yields the correction
factor $G^0_{\rm m}(\omega_0)$ following from the {\it direct} series
expansion:
\beq\label{eq:Gmono_soft_4} 
G^0_{\rm m}(\omega_0)
=1-\frac{21}{5}\,\omega_0+\frac{357}{25}\,\omega_0^2
-\frac{7618}{175}\,\omega_0^3+\frac{21498}{175}\,\omega_0^4
\Abst{.}
\eeq
Figure \ref{fig:Gmono_soft} shows the full numerical result for $G^0_{\rm
m}(\omega_0)$ in comparison to the analytic approximation
\eqref{eq:Gmono_soft_4}, taking into account the corrections up to different
orders in $\omega_0$.
The approximation converges only very slowly and in the highest order
considered here it breaks down close to $\omega_0~\sim~0.15$. Due to this
behavior of the asymptotic expansion one expects no significant improvement
when going to higher orders in $\omega_0$, but the monotonic decrease of the
emission coefficient suggests that a functional form
$G^0_{\rm m}~=~[1+\sum_{k=1}^4\,a_k\,\omega_0^k]^{-1}$
could lead to a better performance. Determining the coefficients $a_i$ by
comparison with the direct expansion \eqref{eq:Gmono_soft_4} one may obtain an
{\it inverse} approximation for the DC correction factor
\beq
\label{eq:G_mono_inv} 
G^{0,\rm inv}_{\rm m}(\omega_0)
\!=\!\frac{1}{1+\frac{21}{5}\,\omega_0+\frac{84}{25}\,\omega_0^2
-\frac{2041}{875}\,\omega_0^3+\frac{9663}{4375}\,\omega_0^4}.
\eeq
As Fig. \ref{fig:Gmono_soft} clearly shows, $G^{0,\rm inv}_{\rm m}(\omega_0)$
provides an excellent description of the numerical result up to very high
energies of the initial photon.
For comparison, we also show the frequency dependence of the total Compton
cross section, $\sigma_{\rm KN}$, as given by the {\it Klein-Nishina} formula
and the correction factor $\kappa(\omega_0)=G^0_{\rm m}/(\sigma_{\rm
KN}/\sigma_{\rm T})$ per act of Compton scattering.
Since it is expected that per {\it single} Compton scattering a small fraction
of electrons and photons undergo DC scattering, this comparison hints towards
the fact that a large part ($\sim 2/3$) of the decrease in the DC emissivity
with $\omega_0$ is due to the decrease in the probability of Compton
scattering photons with larger energies due to quantum-electrodynamic
corrections.
%

\begin{figure}
\centering 
\includegraphics[width=0.96\columnwidth]
{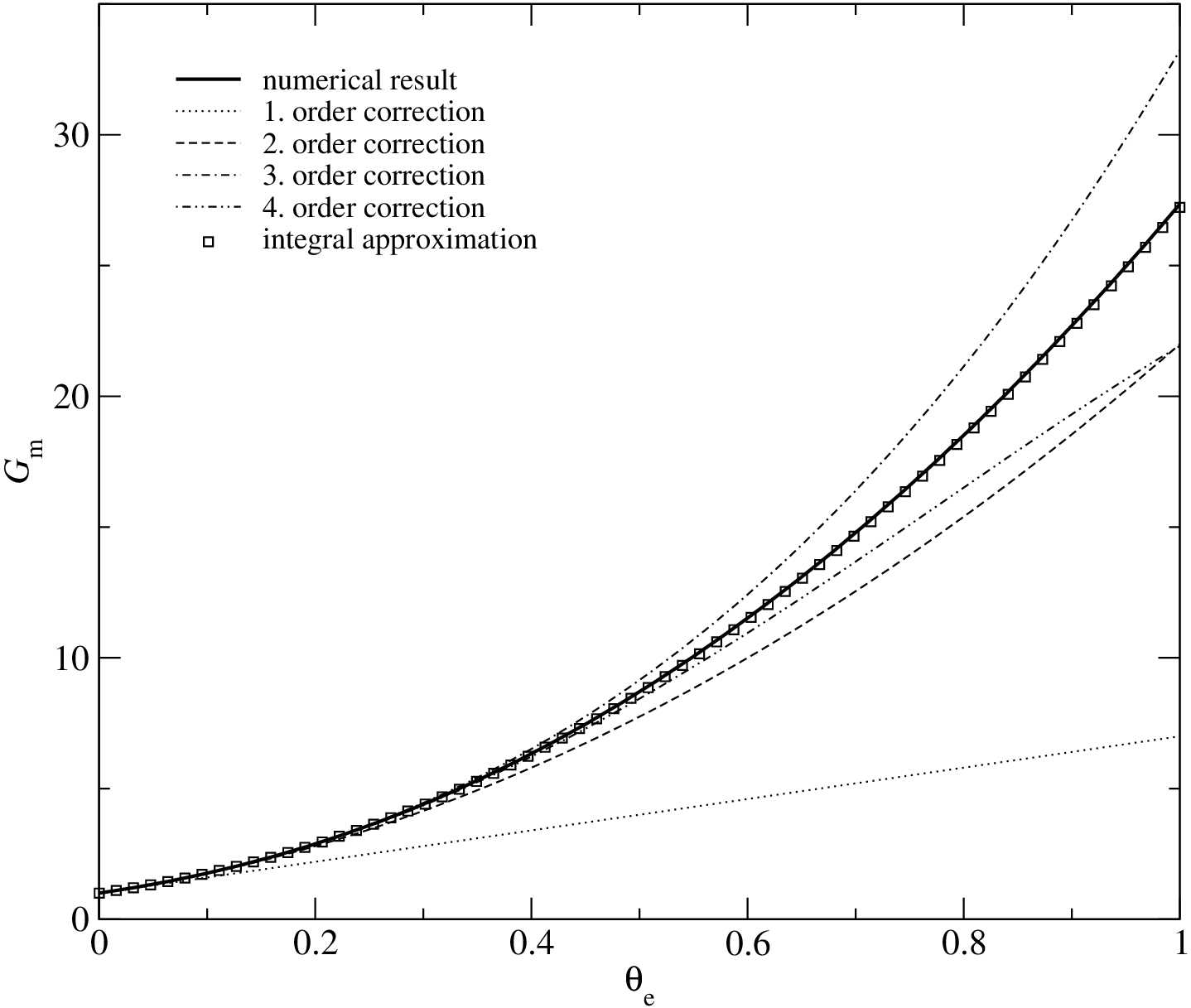}
\caption
{DC emission correction factor $G_{\rm m}^{\rm nr}$ for soft initial photons
  ($\omega_0=10^{-4}$) as a function of the electron temperature $\The=\Te/\me$. Also
  shown is the full integral approximation \eqref{eq:Gmono_soft_nonrel_Ta} and
  the expansion \eqref{eq:Gmono_soft_nonrel_Tb}, taking into account the
  corrections up to different orders in $\The$, as indicated.}
\label{fig:Gmono_soft_nr}
\end{figure}
\subsection{Hot electrons and soft initial photons}
\label{sec:e_nonp}
We now consider the limit $\nu_0\ll \me$ and $\nu_2\ll \nu_0$. Expanding the DC
differential cross section \eqref{eq:dsig_DC} in the lowest orders of $\nu_0$
and $\nu_2$ one obtains
\beal
\label{eq:dsig_DC_soft_non_rel}
\frac{\id \sigDC^{\rm soft}}{\id \Lambda}
&\approx
\frac{\alpha}{2\pi^2}\,\frac{r_0^2}{\gMoei{0}}\,\frac{\omega_0^2}{\nu_2}
\,\left[\gamma^2\lambda_0\lambda_1(\gamma^2\lambda_0\lambda_1-\alpha_{01})+\frac{\alpha_{01}^2}{2}\right]
\nonumber\\
&\;\;\;\; \times
\frac{\gamma^2\lambda_0\lambda_1\lambda^2_2\alpha_{01}-\frac{1}{2}(\lambda_1\alpha_{02}-\lambda_0\alpha_{12})^2}
{\gamma^{10}\lambda_0\lambda_1^6\lambda_2^4} \Abst{.}
\end{align}
with the electron and photon scalars
$\lambda_i=\hat{P}\cdot\hat{K}_i=1-\beta\,\mu_{\rm e\it i}$ and
$\alpha_{ij}=\hat{K}_i\cdot\hat{K}_j=1-\mu_{ij}$.
Now we perform all the angular integrations of the Boltzmann emission integral
\eqref{eq:n2em} but assume monoenergetic electrons, with velocity $\beta_0$.
This then leads to
\beal
\label{eq:n2em_soft_nonrel}
\left.\pAb{n_2}{t}\right|_{\rm em}^{\rm m,nr} = 
\frac{1+\beta_0^2}{1-\beta_0^2}\times \left.\pAb{n_2}{t}\right|_{\rm em,L}^{\rm m}
\Abst{.}
\end{align}
Here the factor of $\gamma_0^2$ can be immediately understood since the
initial photons inside the rest frame of the electron will on average have an
energy of $\omega_{0}^{\rm e}\sim\gamma_0\omega_0$ and hence the DC
emissivity, i.e. $\propto (\omega_{0}^{\rm e})^{2}$, will be $\sim\gamma_0^2$
times larger than for resting electrons.

From Eq.~\eqref{eq:n2em_soft_nonrel} the correction factor for thermal
electrons with temperature $\The$ can be found by averaging over the
relativistic Maxwell-Boltzmann distribution \eqref{eq:relMBD}:
\bsub
\label{eq:Gmono_soft_nonrel_T}
\beal
\label{eq:Gmono_soft_nonrel_Ta} G_{\rm m}^{\rm nr}(\The) 
&= \int_0^\infty
\frac{1+\beta_0^2}{1-\beta_0^2}\,f(E_0)\,p_0^2\id p_0\equiv
2\left<\gamma_0^2\right>_{\rm th}-1
\nonumber\\
&\!\!\!
=
\frac{[1+24\,\The^2]\,K_0(1/\The)+8\,\The\,[1+6\,\The^2]\,K_1(1/\The)}{K_2(1/\The)}
\end{align}
with $E_0=\gamma_0\me$, $p_0=\gamma_0\me\,\beta_0$ and
$\gamma_0=(1-\beta_0^2)^{-1/2}$. 
Here $K_i(1/\The)$ denotes a modified Bessel function of kind $i$. For low
temperatures one can find the approximation
\beq\label{eq:Gmono_soft_nonrel_Tb} 
G_{\rm m}^{\rm nr}(\The)
\stackrel{\stackrel{\The\ll 1}{\downarrow}}{\approx}
1+6\,\theta_{\rm e}+15\,\theta_{\rm e}^2+\frac{45}{4}\,\theta_{\rm e}^3-\frac{45}{4}\,\theta_{\rm e}^4.
\eeq
\esub
Figure \ref{fig:Gmono_soft_nr} shows the numerical result for $G_{\rm m}^{\rm
nr}$ in comparison with the analytic approximations
\eqref{eq:Gmono_soft_nonrel_T}.
The low frequency DC emission strongly increases with temperature. One can
gain a factor of a few for $\The\leq 0.5$ and even a factor $\sim 30$ for
$\The\sim 1$. The expansion \eqref{eq:Gmono_soft_nonrel_Tb} provides an
excellent description of the numerical results up to high temperatures
($\The\sim 0.5$),
but for higher temperatures the integral approximation
\eqref{eq:Gmono_soft_nonrel_Ta} should be used.

\begin{figure}
\centering 
\includegraphics[width=0.96\columnwidth]
{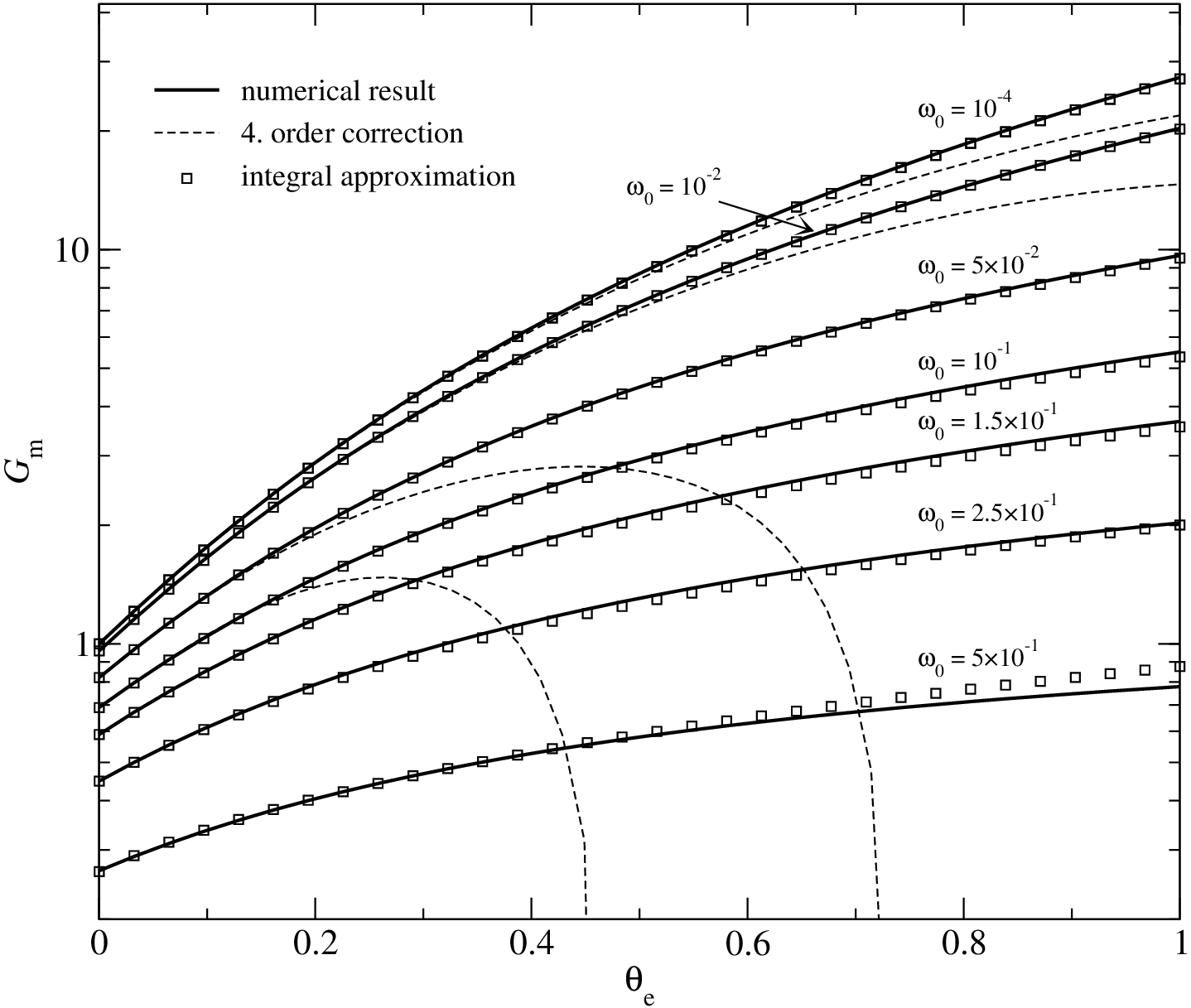}
\caption
{DC emission correction factor $G_{\rm m}$ for different energies of
  the initial photons as a function of $\The=\Te/\me$. Also shown is the full
  integral approximation \eqref{eq:Gmono_soft_w0_T} in combination with the
  inverse factor \eqref{eq:G_mono_inv_beta} and, for some cases, the expansion
  \eqref{eq:Gmono_soft_w0_T_app} taking into account correction terms up to
  fourth order, as indicated.}
\label{fig:DG_Mono_T_w0}
\end{figure}
\subsection{Hot electrons and energetic initial photons}
\label{sec:e_ap}
To improve the analytic description of the DC emission we again return to the
soft photon limit, expanding the DC differential cross section
\eqref{eq:dsig_DC} in the lowest orders of $\nu_2\ll \me$. To include higher
order corrections we also expand this expression up to 4th order in $\omega_0$.
We then carry out all the angular integrations of the Boltzmann emission
integral \eqref{eq:n2em} for monoenergetic electrons, with velocity $\beta_0$.
This leads to
\beal
\label{eq:Gmono_soft_w0_beta} 
G_{\rm m}(\omega_0, \beta_0)
&=
\gamma_0^2
\left[
1+\beta_0^2-\frac{21}{5}\left(1+2\beta_0^2+\frac{1}{5}\beta^4_0\right)\gamma_0\omega_0
\right.
\nonumber\\
&\;\;\left.
+\frac{357}{25}\left(1+\frac{10}{3}\beta_0^2+\beta^4_0\right)\gamma_0^2\omega_0^2
\right.
\nonumber\\
&\;\;\left.
-\frac{7618}{175}\left(1+5\beta_0^2+3\beta^4_0+\frac{1}{7}\beta^6_0\right)\gamma_0^3\omega_0^3
\right.
\nonumber\\
&\;\;\left.
+\frac{21498}{175}\bigg(1+7\beta_0^2+7\beta^4_0+\beta^6_0\bigg)\gamma_0^4\omega_0^4
\right].
\end{align}
As in the previous case the correction factor for thermal electrons with
temperature $\The$ can be found by performing the 1-dimensional integral
\beq\label{eq:Gmono_soft_w0_T} 
\left<G_{\rm m}\right>_{\rm th}
\equiv
G_{\rm m}(\omega_0, \The)
= \int_0^\infty G_{\rm m}(\omega_0, \beta_0)\,f(E_0)\,p_0^2\id p_0
\Abst{.}
\eeq
Unfortunately here no full solution in terms of simple elementary functions
can be given, but the integral can be easily evaluated numerically. 
In the limit of low temperatures one can find the simple approximation
\beal
\label{eq:Gmono_soft_w0_T_app} 
G_{\rm m}(\omega_0, \The)
&\approx
1-\frac{21}{5}\,\omega_0+\frac{357}{25}\,{\omega}_0^2-\frac{7618}{175}\,{\omega}_0^3+\frac{21498}{175}\,{\omega}_0^4
\nonumber\\
&
\quad\quad
+\left[6- \frac{441}{10}\,{\omega}_0+ \frac{5712}{25}\,{\omega}_0^2-\frac{34281}{35}\,{\omega}_0^3\right]\The
\nonumber\\
&
\quad\quad\;\;+
\left[15 - \frac{8379}{40}\,{\omega}_0+ \frac{8568}{5}\,{\omega}_0^2\right]\The^2 
\nonumber\\
&
\quad\quad\;\;\;\;\,+
\left[\frac{45}{4}- \frac{3969}{8}\,{\omega}_0\right]\The^3 -\frac{45}{4}\,\The^4 
\end{align}
for \eqref{eq:Gmono_soft_w0_T}, which clearly shows the connection to the
two limits discussed above, i.e. Eqn. \eqref{eq:Gmono_soft_4} and
\eqref{eq:Gmono_soft_nonrel_Tb}.

As was shown in Sect. \ref{sec:ce_ap} an {\it inverse} Ansatz for $G_{\rm m}$
led to an excellent approximation for the full numerical result in the cold
plasma case. To improve the performance of the obtained analytic
approximation for $G_{\rm m}(\omega_0, \beta_0)$ we tried many different
functional forms, always comparing with the direct expansion
\eqref{eq:Gmono_soft_w0_beta}.  After many attempts we found that
\bsub
\label{eq:G_mono_inv_beta_gen} 
\beq
\label{eq:G_mono_inv_beta} 
G^{\rm inv}_{\rm m}(\omega_0, \beta_0)
= \frac{\gamma_0^2 \,(1+\beta_0^2)}{1+\sum_{k=1}^4 f_k(\beta_0)\,\gamma_0^k\omega_0^k}
\Abst{,}
\eeq
with the functions $f_k(\beta_0)$
\beal
\label{eq:Gmono_soft_w0_beta_inv} 
f_1(\beta_0)
&=\;\;\,\frac{1}{1+\beta_0^2}\,\left[\frac{21}{5}+\frac{42}{5}\beta_0^2+\frac{21}{25}\beta^4_0\right]
\\
f_2(\beta_0)
&=\;\;\,\frac{1}{(1+\beta_0^2)^2}\,\left[\frac{84}{25}+\frac{217}{25}\beta_0^2+\frac{1967}{125}\beta^4_0\right]
\\
f_3(\beta_0)
&=-\frac{1}{(1+\beta_0^2)^3}\,\left[\frac{2041}{25}+\frac{1306}{125}\beta_0^2\right]
\\
f_4(\beta_0)
&=\;\;\,\frac{1}{(1+\beta_0^2)^4}\,\frac{9663}{4375}
\end{align}
\esub
provides the best description of the full numerical results.

Figure \ref{fig:DG_Mono_T_w0} shows the numerical results for $G_{\rm
  m}(\omega_0, \The)$ in comparison with the integral approximation
\eqref{eq:Gmono_soft_w0_T} in combination with the inverse factor
\eqref{eq:G_mono_inv_beta} and the direct expansion
\eqref{eq:Gmono_soft_w0_T_app} taking into account correction terms up to
fourth order.
The performance of the integral approximation is excellent in the full range
of considered cases, but the direct expansion breaks down at lower and lower
temperatures once the initial photon frequency increases. For $\omega_0\leq
0.1$ the direct expansion should be applicable within a few percent accuracy
up to $\The\sim 0.2$.

\begin{figure}
\centering 
\includegraphics[width=0.96\columnwidth]
{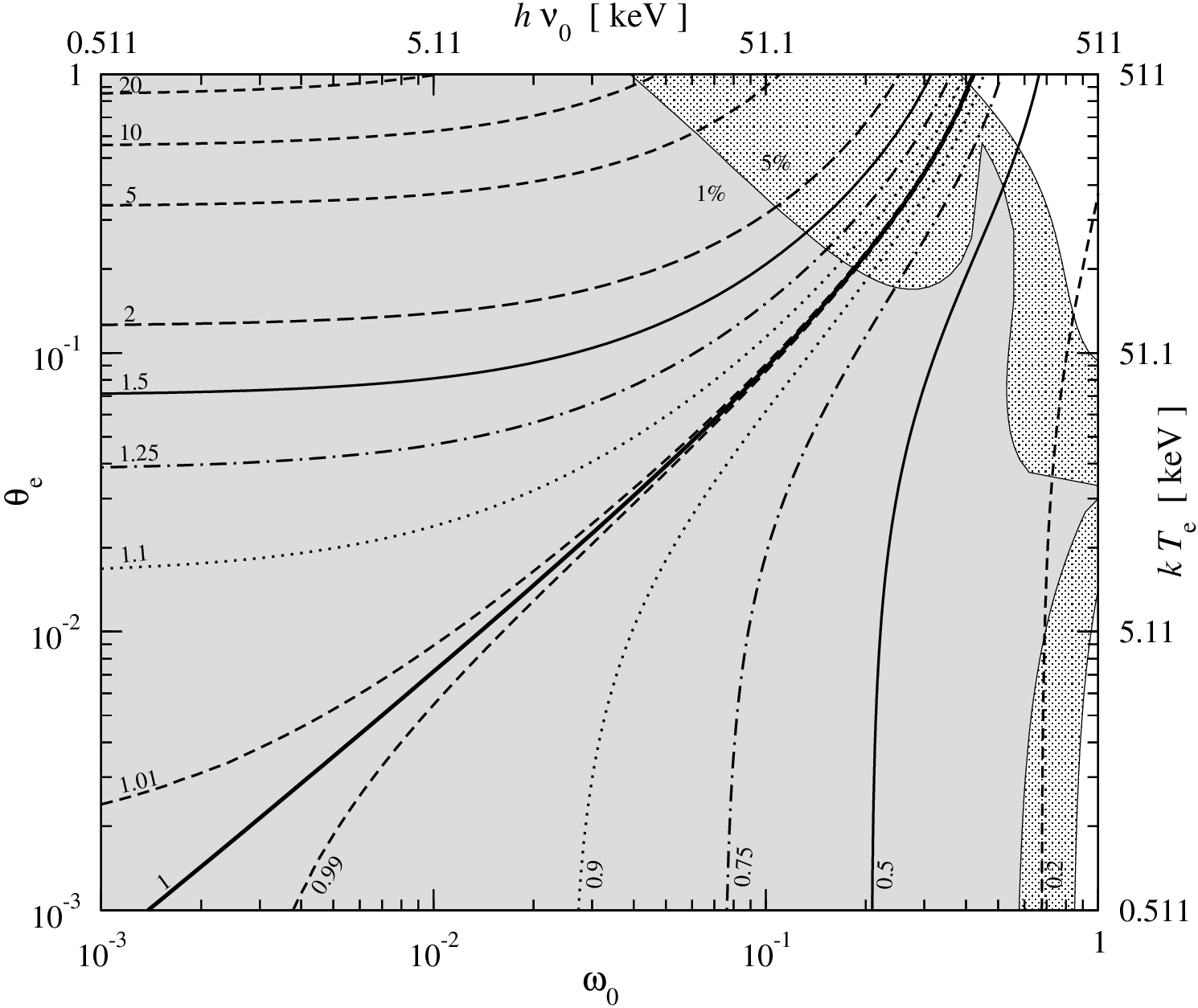}
\caption
{DC emission correction factor for monoenergetic initial photons and thermal
  electrons. The lines indicate the contours of $G_{\rm m}=\text{const}$ as
  obtained by full numerical integrations. Within the shaded regions the
  approximation based on \eqref{eq:G_mono_inv_beta_gen} is accurate to better
  than 1\% and 5\%, respectively.}
\label{fig:Rashid_mono}
\end{figure}
In Fig.\ref{fig:Rashid_mono} the dependence of the DC emission correction
factor on the energy of the initial photons and the temperature of the
electrons is illustrated. From this one can see that the Lightman-Thorne
approximation is applicable in a very limited range of photon energies and
electron temperatures. On the other hand the approximation based on
\eqref{eq:G_mono_inv_beta_gen} works in practically the full considered range
with very high accuracy.
Focusing on the curve $G_{\rm m}=1$ one can see that for low energies of the
initial photon $(\omega_0\lesssim 10^{-2})$ the required electron temperature
scales like $\The\sim 0.7\,\omega_0$. Considering the first order correction
terms in Eq.~\eqref{eq:Gmono_soft_w0_T_app} one also finds this scaling. Up to
$\omega_0\gtrsim 0.15$ the necessary temperature of the electrons is
$\The<\omega_0$. This shows that moving electrons can easily compensate the
suppression of DC emission due to the increasing energy of the initial
photon. However, for $\omega_0\gtrsim 0.15$ the electrons have to be up to
several times hotter to accomplish this.

%
%
%
\section{DC emission for more general initial photon distributions}
Based on the results given in the previous Section one can obtain simple
expressions for more general initial photon distributions. 
For initial photons with phase space distribution $n(\nu)$ within the {\it
Lightman-Thorne approximation} one finds
\beal
\label{eq:BRform}
\left.\pAb{n_2}{t}\right|_{\rm DC,\,em}^{\rm soft} &=
\frac{4\alpha}{3\pi}\,\Ne\,\sigT\,
\frac{\Thg^2}{x^3_{2}}\times I_0
\Abst{,}
\end{align}
where here we defined the dimensionless photon frequency
$x_{i}=\omega_i/\Thg=\nu_i/\Tg$. \change{We also introduced the {\it effective
    temperature} $\Tg$ which characterizes the total energy density of the
  photon field. For Planckian photons $\Tg$ is identical with the {\it
    thermodynamic temperature}.}

The DC emission factor $I_0$ is given by the integral (compare
\citet{Light81}, Eq. (10b) for $n(\nu)\ll1$):
\beal
\label{eq:G_0}
I_0&=\int^\infty_0 x^4\,n(x)\id x
\Abst{.}
\end{align}
over the initial photon distribution $n(\nu)$. Stimulated DC emission was
neglected, \change{but we will discuss possible extensions in
  Sect.~\ref{sec:induced}}.
Here one factor of $x^2$ arises from the conversion $N_0 \rightarrow
\nu^2 n(\nu)$, \change{and} the other is due to the scaling of the {\it Lightman-Thorne
approximation} Eq.~\eqref{eq:n2em_Light} with $\omega_0$.
Note that here and below we omit the index '0' for the \change{frequency of
  the} initial photon.

Now including higher order corrections in the energies of the initial photon
and electrons, but still in lowest order of $x_2$, instead of $I_0$ one will
obtain a more general function $I$. With this we can define an {\it effective
  DC emission correction factor} in the soft photon limit by
\beal
\label{eq:Gaunt_DC}
g_{\rm dc}^{\rm soft}=I/I_0.
\end{align}
We shall now give analytic expressions for $I$
based on the results obtained in the previous Section.

\subsection{The effective DC emission correction factor}
Here we will give analytic expressions for the effective DC emission
correction factor with {\it three different} approaches, \change{each with
  their advantages and disadvantages}. We will discuss the range of
applicability for these expressions for Wien spectra in the next Section.

\subsubsection{Approximation based on $G^{\rm inv}_{\rm m}$} 
As was shown in Section \ref{sec:e_ap} the expression
\eqref{eq:G_mono_inv_beta_gen} averaged over a thermal electron distribution
(compare with Eq.~\eqref{eq:Gmono_soft_w0_T}) provides an excellent
description of the DC emissivity for monochromatic initial photons. If induced
effects are negligible one can simply convolve this result with the
corresponding initial photon distribution $n(\nu)$ to obtain:
\beq
\label{eq:Gmono_inv} 
I= \int_0^\infty x^4 \,n(x) \left<G^{\rm inv}_{\rm m}\right>_{\rm th}\id x
\Abst{.}
\eeq
Here $\left<G^{\rm inv}_{\rm m}\right>_{\rm th}$ denotes the thermally
averaged expression~\eqref{eq:G_mono_inv_beta_gen}. 
As will be shown below, the approximation \eqref{eq:Gmono_inv} works very well
in a broad range of different temperatures. However, it involves a
2-dimensional integral which in numerical applications may be too demanding.

\subsubsection{Direct expansion of $G_{\rm m}$} 
One can also replace $\left<G^{\rm inv}_{\rm m}\right>_{\rm th}$ in Eq.
\eqref{eq:Gmono_inv} with the more simple analytic expression
\eqref{eq:Gmono_soft_w0_T_app} derived in the limit of low electron
temperature and energy of the initial photon. In this case the DC emission
coefficient $I$ can be written as the {\it direct} expansion
\beal\label{eq:Gfunctionres}
I^{\rm exp}=\sum^4_{k=0} I_k \times\theta_{\rm e}^k
\end{align}
where the definitions of the integrals $I_i$ may be found in Appendix
\eqref{eq:GFunc}. The first integral $I_0$ corresponds to the result obtained
by \citet{Light81} and \citet{Thorne81} in the limit $n(\nu)\ll 1$.
\change{However, it is expected that this approximation in general converges
  very slowly and is mainly useful for simple analytic estimates.}

\subsubsection{Alternative approach} 
\label{sec:Beyond}
Above we gave the expressions obtained from the {\it direct} expansion of $I$.
However, this procedure results in an asymptotic expansion of the effective DC
correction factor $g_{\rm dc}^{\rm soft}$ or equivalently $I$, which in most
of the cases is expected to converge very slowly.
Therefore, using the {\it direct} expansion \eqref{eq:Gfunctionres} we again
tried to `guess' the correct functional form of the DC emission coefficient
$I$ and thereby to extend the applicability of this simple analytic
expression.
The general behavior of the results obtained in our numerical integrations
showed that for $\theta_{\rm e}\lesssim\theta_{\gamma}$ the effective DC
correction factor $g_{\rm dc}^{\rm soft}=I/I_0$ decreases towards higher
electron temperatures.  After several attempts we found that this behavior is
best represented by the functional form
\bsub
\label{eq:Ggeneralinv} 
\beq
\label{eq:Ggenerala} 
I^{\rm inv}=\frac{I_0}{1+a\,\theta_{\rm e}}
\Abst{.}
\eeq
Using Eq. \eqref{eq:Gfunctionres}, one finds
\beq 
\label{eq:coeffa}
a=\frac{21}{5}\,\frac{\Thg}{\The}\,\frac{\int x^5 \,n \id x}{\int x^4\,n\id x}-6
\Abst{.}
\eeq
\esub
As will be shown below, equation \eqref{eq:Ggenerala} together with
\eqref{eq:coeffa} provides a description of the low frequency DC emission
coefficient, which for photon distributions close to a Wien spectrum is
accurate to better than \err\% for temperatures up to $\sim 25$ keV in the
range $0.2\leq\rho\leq 50$. Here we defined the ratio of the electron to
photon temperature as $\rho=\Te/\Tg$. For a detailed discussion we refer the
reader to Sect.  \ref{sec:Dis}.
Since this approximation involves only 1-dimensional integrals over the initial
photon distribution in numerical applications it may be more useful than the
approximation \eqref{eq:Gmono_inv}.

\subsection{Approximate inclusion of stimulated DC emission}
\label{sec:induced}
\change{ For broad initial photon distributions, stimulated DC emission arises
  which, in particular for situations close to full thermodynamic equilibrium,
  can become significant. Accounting for this effect, the DC emission integral
  contains additional factors of $1+n(\nu_1)$ and $1+n(\nu_2)$ due to the
  presence of ambient photons at frequencies $\nu_1$ and $\nu_2$. If we are
  interested in the emission of photons at frequency $\nu_2$ then $1+n(\nu_2)$
  appears as a global factor in front of the full Boltzmann emission integral
  and no further complications arise due to this term. However, the factor
  $1+n(\nu_1)$ remains inside the integrand and the dependence of the
  scattered photon frequency on the scattering angles and energies has to be
  taken into account.

  Within the Lightman-Thorne approximation, for a particular DC scattering
  event one always has $\nu_0\sim\nu_1$, since for $\Te=0$ no {\it Doppler
    boosting} appears and for $\nu_0\ll \me$ the {\it recoil effect} is
  negligible. Therefore one can write
\beal
\label{eq:G_0_stim}
I^{\rm stim}_0&\approx[1+n(x_2)]\times\int^\infty_0 x^4\,n(x)[1+n(x)]\id x
\end{align}
instead of Eq. \eqref{eq:G_0}. In situations close to full thermodynamic
equilibrium the main effect due to stimulated DC emission is {\it only} due to
the factor $1+n(x_2)$, whereas the correction related to $1+n(x)$ inside the
integral is only of a few percent. Assuming $n(x)=1/[e^x-1]$ yields $I_0\sim
24.89$ and $I^{\rm stim}_0\sim 25.98/x_2$ for $x_2\ll 1$ which further
supports this statement.

Similarly, one can modify Eq.~\eqref{eq:Gmono_inv} with the replacement
$n(x)\rightarrow n(x)[1+n(x)][1+n(x_2)]$, but again this approach is limited
to cases when the change of the initial photon energy is very small
($\nu_0\sim \nu_1$) and the radiation spectrum is broad ($\Delta\nu/\nu \gg
\beta$).
This then yields
\beal
\label{eq:G_stim}
I^{\rm stim}&\approx[1+n(x_2)]\times\int^\infty_0 x^4\,n(x)[1+n(x)]\left<G^{\rm inv}_{\rm m}\right>_{\rm th}\id x
\end{align}
Now, looking at the ratio 
$g_{\rm dc}^{\rm stim}=I^{\rm stim}/I^{\rm stim}_0$
one again expects that due to stimulated processes the effective DC emission
correction factor will also be affected at the percent-level only. We
confirmed this statement for Planckian initial photons with $\Tg=\Te$
performing the full Boltzmann emission integral, but a more detailed treatment
of this problem is beyond the scope of this paper, and as mentioned above
below we restrict ourselves to cases in which stimulated DC emission is
negligible.
Furthermore it is clear that in situations when stimulated DC emission becomes
important, DC absorption also should be taken into account and a full
derivation of the kinetic equation for the evolution of the photon field under
the DC process is required.
In lowest order and for situations very close to full thermodynamic
equilibrium, accounting for higher order corrections in the initial photon
energy and electron temperature, one can simply multiply the {\it kinetic
  equation} for the time evolution of the photon field under DC scattering as
formulated by \citet{Light81} and \citet{Thorne81}, by the effective DC
correction factor $g_{\rm dc}^{\rm soft}$, Eq. \eqref{eq:Gaunt_DC}, or
alternatively $g_{\rm dc}^{\rm stim}=I^{\rm stim}/I^{\rm stim}_0$.
However, in general the situation is much more complicated and a detail
treatment of the full Boltzmann collision term is required, which will be left
for a future publication.  
}

\section{Results for Wien spectra}
\label{sec:nWien}
\subsection{Analytic expressions}
For a Wien spectrum $\nW(x)=N_0\,e^{-x}$ of temperature $\Thg$, with the
condition $N_0\ll \nu^2$, induced terms may be neglected.
Above $N_0$ in physical units is defined by $N_{\gamma, 0}=8\pi\, N_0/c^3$,
where $N_{\gamma, 0}$ is the photon number density.
Then the functions $I_i$ are given by
\bsub\label{eq:GWien}
\beal 
I_{0,\rm W}&= \;\;\,24\,N_0
\\
\label{eq:GWien_b}
I_{1,\rm W}&= -I_{0,\rm W}[21 - 6\,\rho]/\rho
\\
I_{2,\rm W}&= \;\;\,I_{0,\rm W}
[428.4-220.5\,\rho + 15\,{\rho }^2]/\rho^2
\\
I_{3,\rm W}&=-I_{0,\rm W}
[9141.6 - 6854.4\,\rho 
\nonumber\\
&\qquad\qquad\quad
+ 1047.375\,{\rho }^2 - 11.25\,{\rho }^3]/\rho^3
\\
I_{4,\rm W}&=\;\;\,I_{0,\rm W}
[206380.8-205686\,\rho  + 51408\,{\rho }^2 
\nonumber\\
&\qquad\qquad\quad
- 2480.625\,{\rho}^3 - 11.25\,{\rho }^4]/\rho^4 
\Abst{.}
\end{align}
\esub
Here we defined the ratio of the electron to photon temperature as
$\rho=\Te/\Tg$.  
Making use of \eqref{eq:Ggenerala} and \eqref{eq:coeffa} one finds:
\beq\label{eq:GWienrho} 
I^{\rm inv}_{\rm W}
=
\frac{24\,N_0}{1 + \left[ 21 - 6\,\rho\right]\,\theta_{\gamma}}
\Abst{.}
\eeq
Here we replaced the electron temperature by $\theta_{\rm
e}\rightarrow\rho\,\theta_{\gamma}$.

\begin{figure}
\centering 
\includegraphics[width=0.96\columnwidth]
{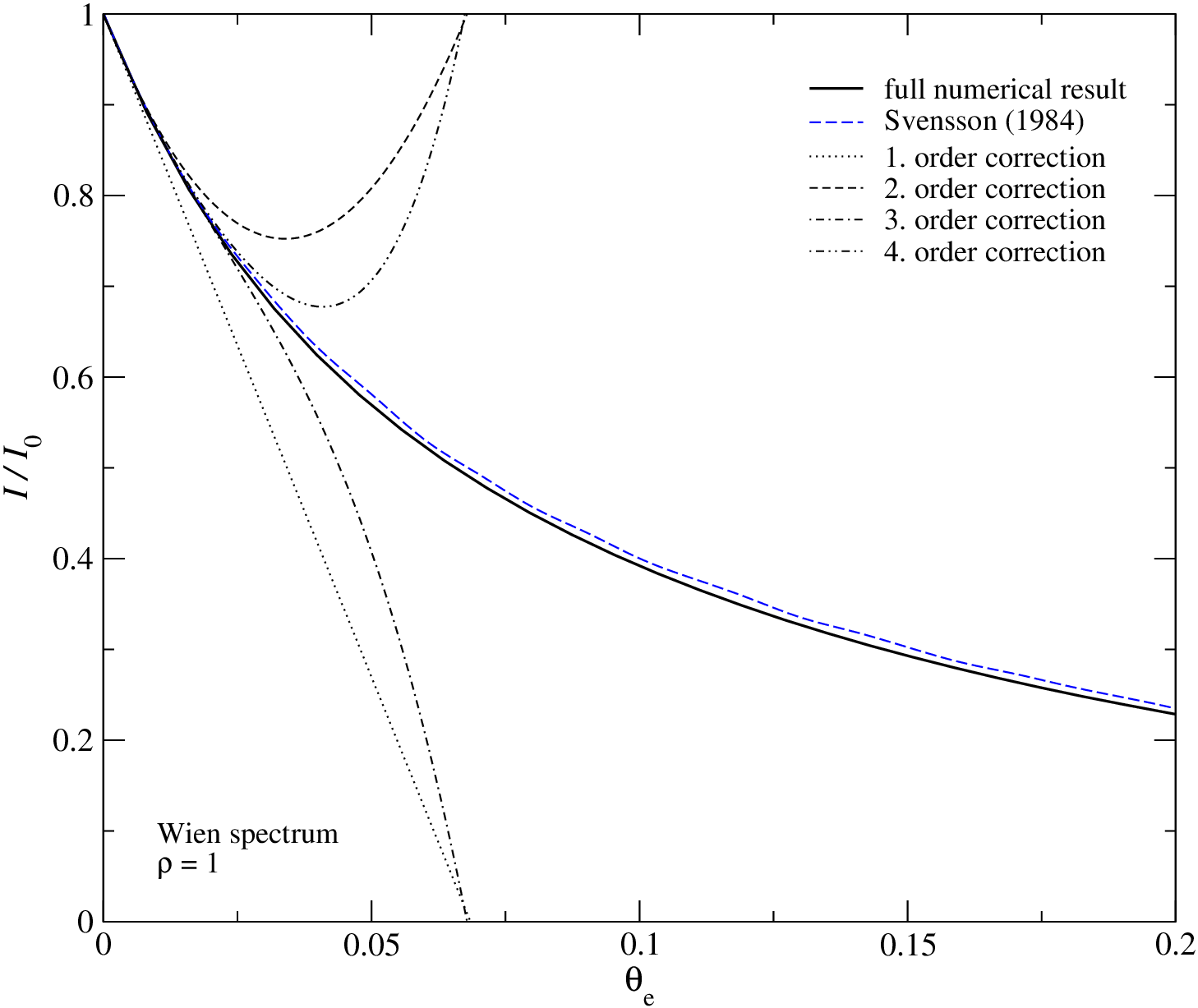}
\caption
{DC soft photon production rate relative to the DC production rate $I_0$ as
  given by the Lightman-Thorne approximation as a function of the electron
  temperature $\The=\Te/\me$. All the curves were computed for initial
  photons with a Wien spectrum at temperature $\Tg\equiv\Te$. The full
  numerical results are shown in comparison with the {\it direct} expansion as
  given by equations \eqref{eq:GWien} including temperature corrections up to
  different orders, as indicated.
  In addition, the approximation formula \eqref{eq:Svensson} as given by
  \citet{Svensson84} is shown.  The approximations based on formulae
  \eqref{eq:Gmono_inv} and \eqref{eq:GWienrho} show a similarly good
  performance but for clarity are not presented here. For further discussion
  about the performance of the various approximations see Sect.~\ref{sec:app}.
}
\label{spectra}
\end{figure}
\subsection{Comparison with numerical results}
\label{sec:Dis}

\subsubsection{Case $\Te=\Tg$}
In Fig.~\ref{spectra} the results obtained for incoming Wien spectra with
\change{effective} temperature $\Thg=\theta_{\rm e}$ are shown.
With increasing temperature the effective DC correction factor decreases strongly
in both cases. 
This implies that for higher temperatures the efficiency of DC emission is
significantly overestimated by the result obtained by \citet{Light81} and
\citet{Thorne81}. For example, even at moderate temperatures $\Te\sim
4\,\text{keV}$ there is a $10\%$ negative correction due to higher order
corrections in the energies of the initial photons and electrons.
Since $I/I_0<1$, with Eq.~\eqref{eq:Gmono_soft_w0_T_app} one can estimate that
the main contribution to $I$ has to come from photons with energies
$\omega_0\gtrsim \frac{10}{7}\,\The$.

In the considered case the direct formulae \eqref{eq:Gfunctionres} in fourth
order of the electron temperature with \eqref{eq:GWien} are valid up to
$\theta_{\rm e}\sim 0.05$, within reasonable accuracy. As mentioned above the
convergence of these asymptotic expansions is very slow.
However, as will be shown below, the inverse formula \eqref{eq:Ggenerala} and
the corresponding coefficients $a$ for $\rho\sim 1$ provide an approximation
which is better than \err\% up to $\theta_{\rm e}\sim 0.2$.
Comparing the result for an incoming Wien spectrum with the approximation
given by \citet{Svensson84},
\beal
\label{eq:Svensson}
I_{\rm S}(\The)\approx\frac{I_{0,\rm W}}{1+13.91\,\The+11.05\,\The^2+19.92\,\The^3}
\end{align}
shows that for $\rho=1$ Eq.~\eqref{eq:Svensson} provides a very good fit to
the full numerical results. The approximations based on formulae
\eqref{eq:Gmono_inv} and \eqref{eq:GWienrho} show a similarly good
performance. However, in the more general case, i.e. $\rho\neq 1$, the
approximation given by \citet{Svensson84} does not reproduce the numerical
results, since the strong dependence of the higher order corrections on the
ratio of the electron to the photon temperature was not taken into account.

\begin{figure}
\centering 
\includegraphics[width=0.96\columnwidth]
{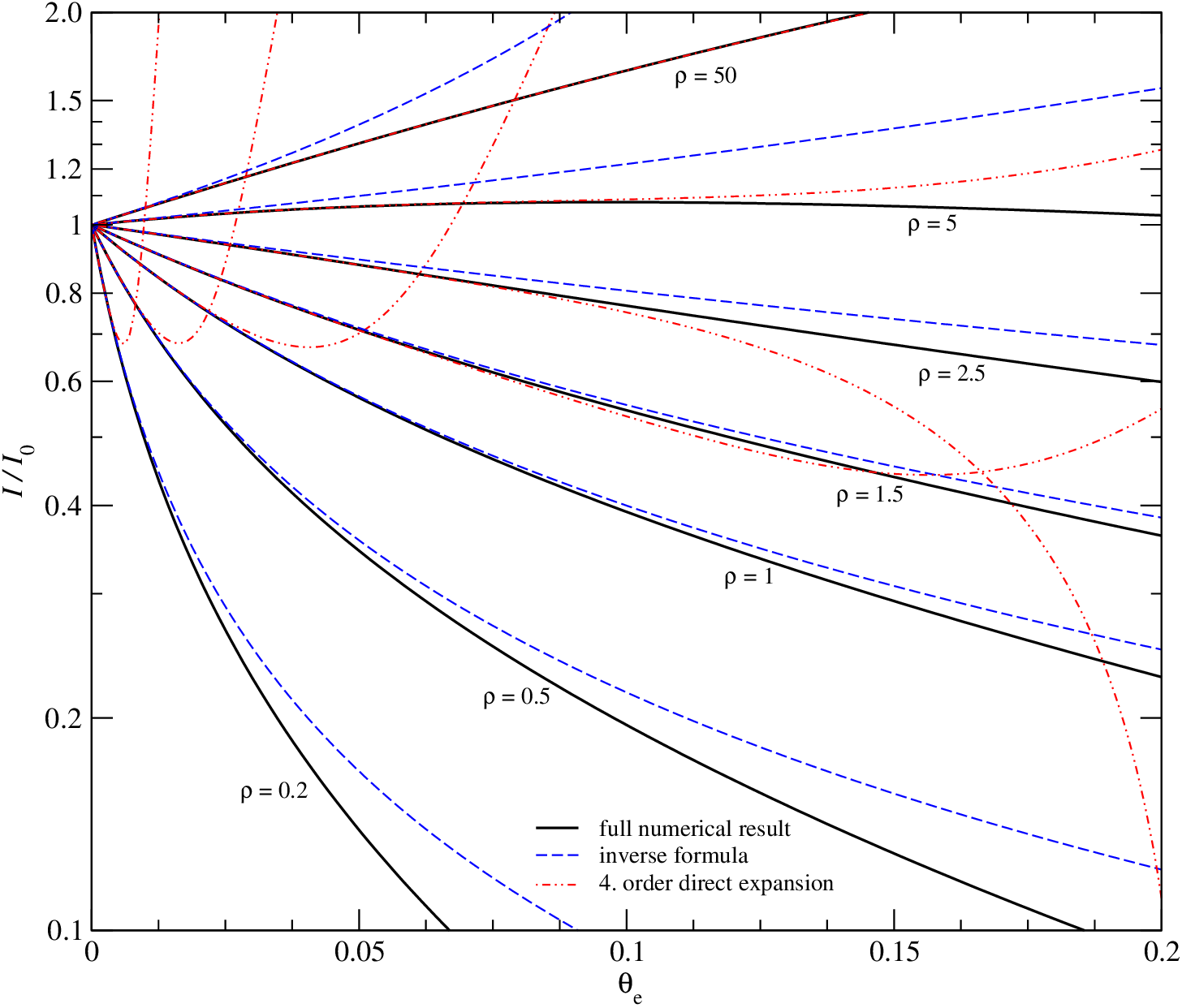}
\caption
{DC soft photon production rate relative to the DC production rate $I_0$ as
  given by the Lightman-Thorne approximation as a function of the electron
  temperature $\The=\Te/\me$. All the curves were computed for initial
  photons with a Wien spectrum at temperature $\Tg$ and different values of
  $\rho=\Te/\Tg$.
  The full numerical result, the fourth order direct formula as given by
  Eq.~\eqref{eq:GWien} and the inverse formula from Eq.~\eqref{eq:GWienrho}
  are given.}
\label{Wien_rho}
\end{figure}
\subsubsection{Case $\Te\neq\Tg$}
Figure \ref{Wien_rho} illustrates the dependence of the effective DC
correction factor on the ratio of the electron to the photon temperature
$\rho=\Te/\Tg$ for an initial Wien spectra. 
In general the characteristics of the ratio $I/I_0$ can be summarized as
follows: If we define the critical ratio $\rho_{\rm c}=7/2\sim 2.33$ of
electron to photon temperatures, for which the first order correction to $I$
vanishes (cf. Eq. \eqref{eq:GWien_b}), then two regimes can be distinguished:
(i) for $\rho\lesssim \rho_{\rm c}$ the ratio $I/I_0$ monotonically decreases
with increasing electron temperature, whereas (ii) for $\rho\gtrsim\rho_{\rm
  c}$ it first increases and then turns into a decrease at high temperature.
This suggests that $\rho_{\rm c}$ separates the regions where Doppler boosting
is compensating the suppression of DC emission due to higher photon energy.
Estimating the mean photon energy weighted by $I_0$, i.e.
$\left<\omega_0\right>=\Thg\,\int x^5 n\id x/\int x^4 n\id x \approx 5\,\Thg$,
and using $\omega_0\sim\frac{10}{7}\,\The$ for the condition $G_{\rm
  m}(\omega_0, \The)=1$ (see Eq.~\eqref{eq:Gmono_soft_w0_T_app}) yields
$\rho_{\rm c}\approx 7/2$, which further supports this conclusion.

In general one finds (see Figs. \ref{Wien_rho}) that for $\rho<\rho_{\rm c}$
the {\it inverse} approximation \eqref{eq:Ggenerala} has a better performance
than the {\it direct} expansion formula \eqref{eq:Gfunctionres} and vice versa
for the case $\rho\geq\rho_{\rm c}$. The direct formula performs very well for
$\rho\geq\rho_{\rm c}$ even in lower orders of the temperature corrections.
Combining these two approximations an accuracy of better than $\sim5\%$ can be
achieved over a very broad range of temperatures.  If the photons and
electrons have similar temperatures, then the inverse formula
\eqref{eq:Ggenerala} is valid even up to $\kB T\sim 100\,$keV. In
Section~\ref{sec:app} we discuss the range of applicability for the
various approximations given above in more detail.

\begin{figure}
\centering 
\includegraphics[width=0.96\columnwidth]
{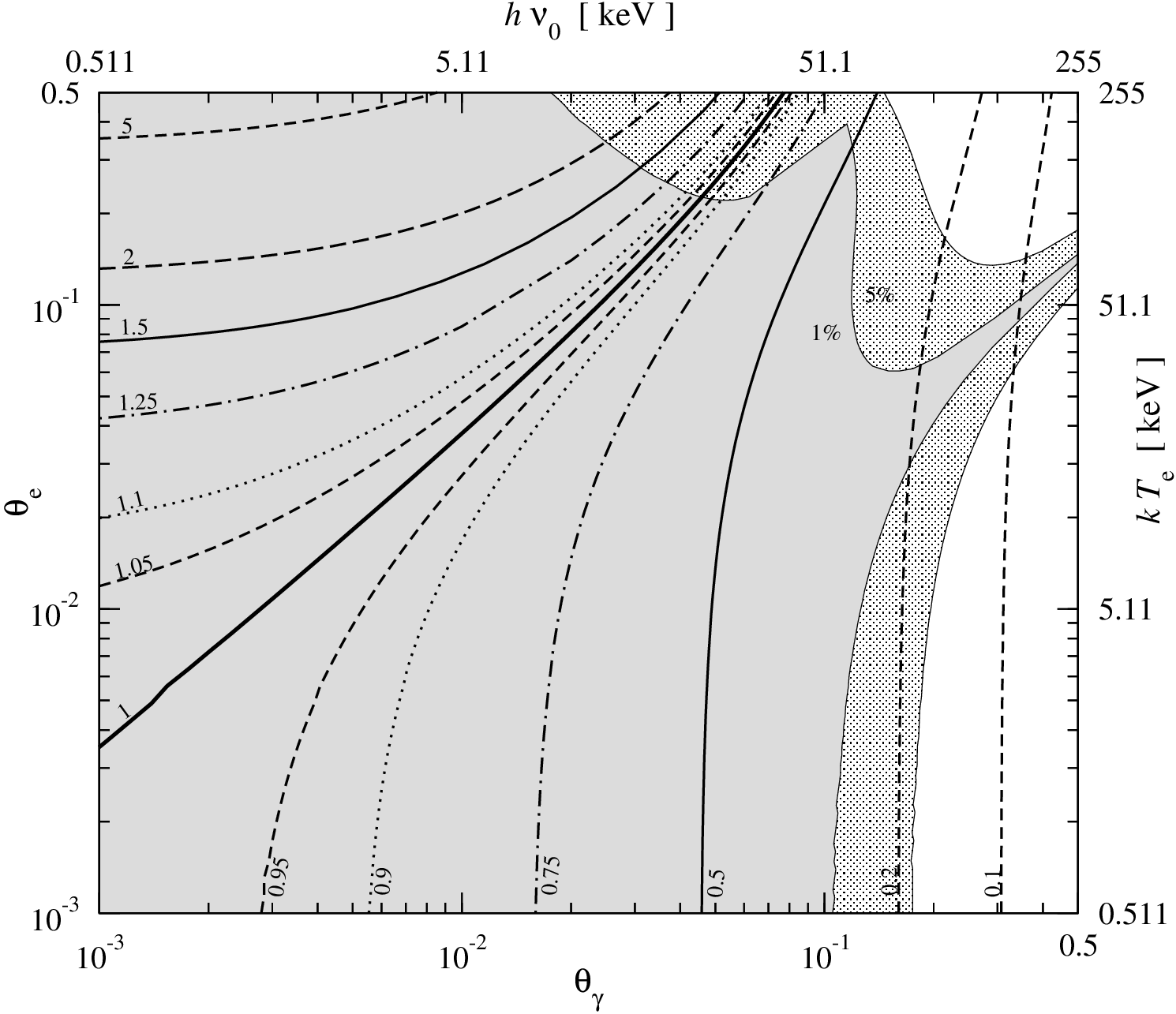}
\caption
{DC emission correction factor for Wien spectra. The lines indicate the
contours of $I/I_0=\text{const}$ as obtained by full numerical
integrations. Within the shaded regions the approximation~\eqref{eq:Gmono_inv}
is accurate to better than 1\% and 5\%, as indicated.
}
\label{fig:Rashid_Wien}
\end{figure}
In Fig.\ref{fig:Rashid_Wien} the dependence of the DC emission correction
factor on the temperature of the initial photons and the temperature of the
electrons is illustrated for Wien spectra. 
The Lightman-Thorne approximation is accurate in a very limited range of
photon energies and electron temperatures. On the other hand, the
approximation~\eqref{eq:Gmono_inv} works in practically the full considered
range with very high accuracy.
In contrast to the case of monoenergetic initial photons even at low
temperatures of the photons the electrons have to be several times hotter in
order to reach $I/I_0=1$. As discussed above this is due to the fact that the
main emission is coming from photons with $\omega_0>\text{few}\times \Thg$.

\begin{figure}
\centering 
\includegraphics[width=0.96\columnwidth, height=0.70\columnwidth]{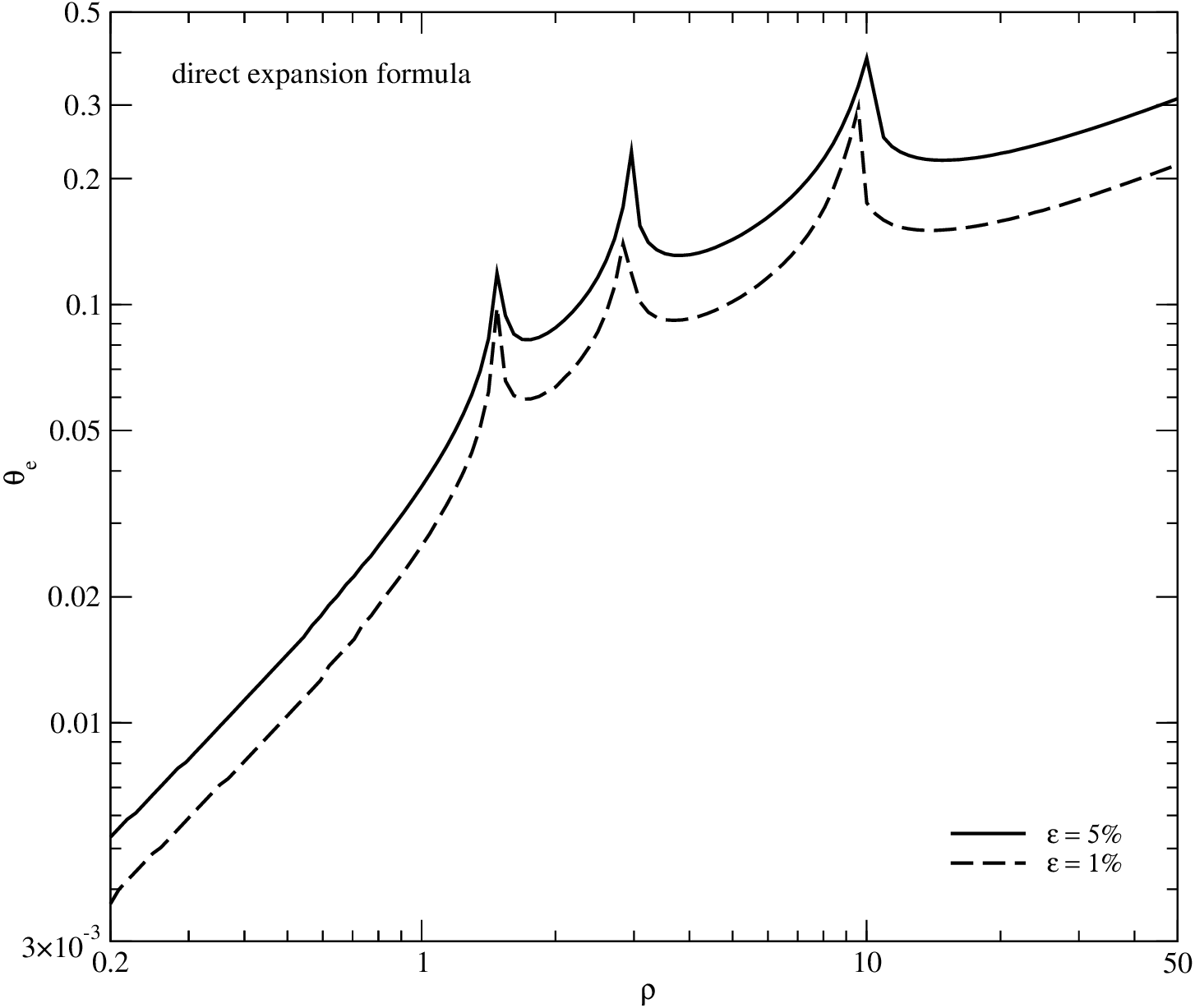}
\\
\includegraphics[width=0.96\columnwidth, height=0.70\columnwidth]{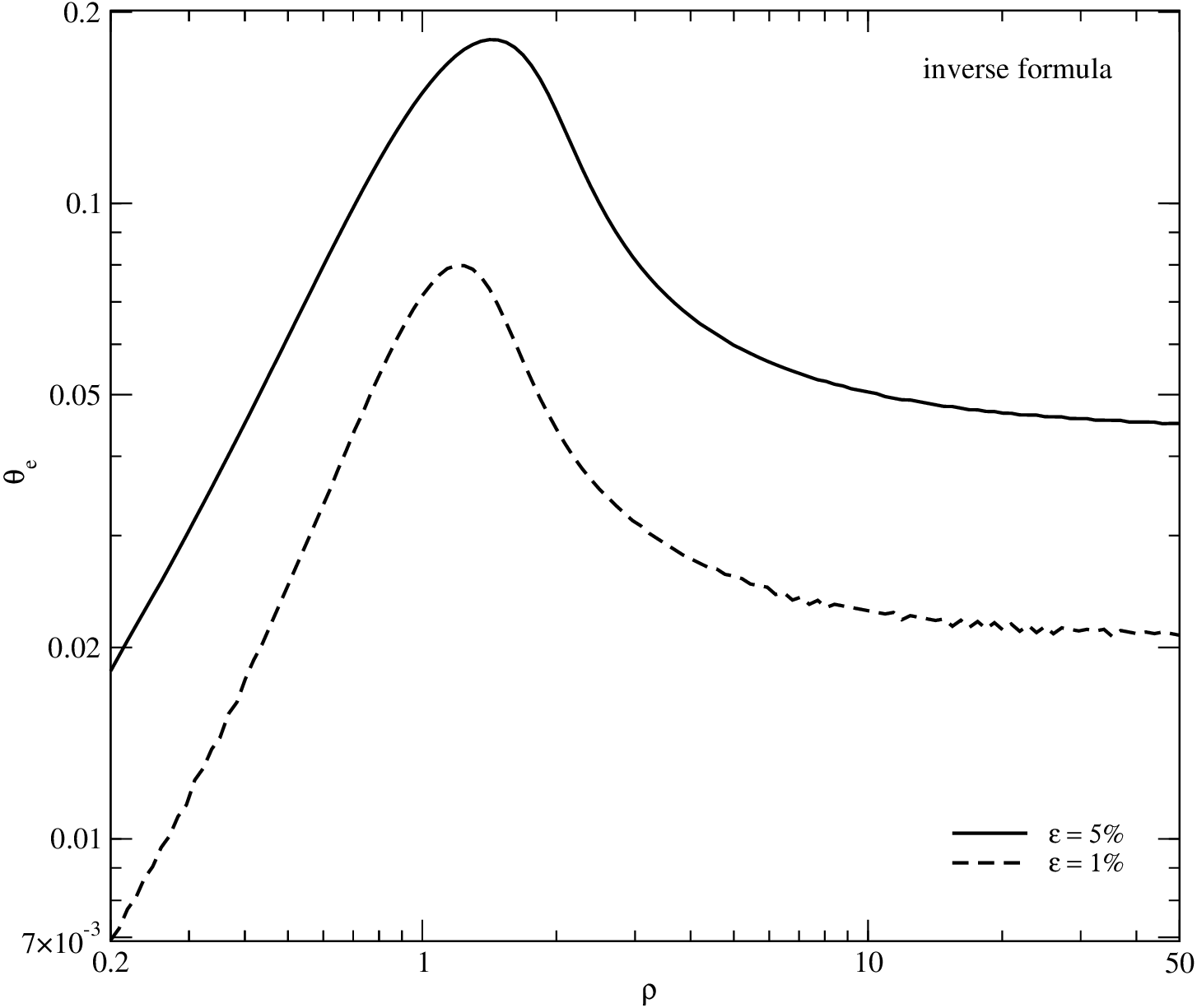}
\\
\includegraphics[width=0.96\columnwidth, height=0.70\columnwidth]{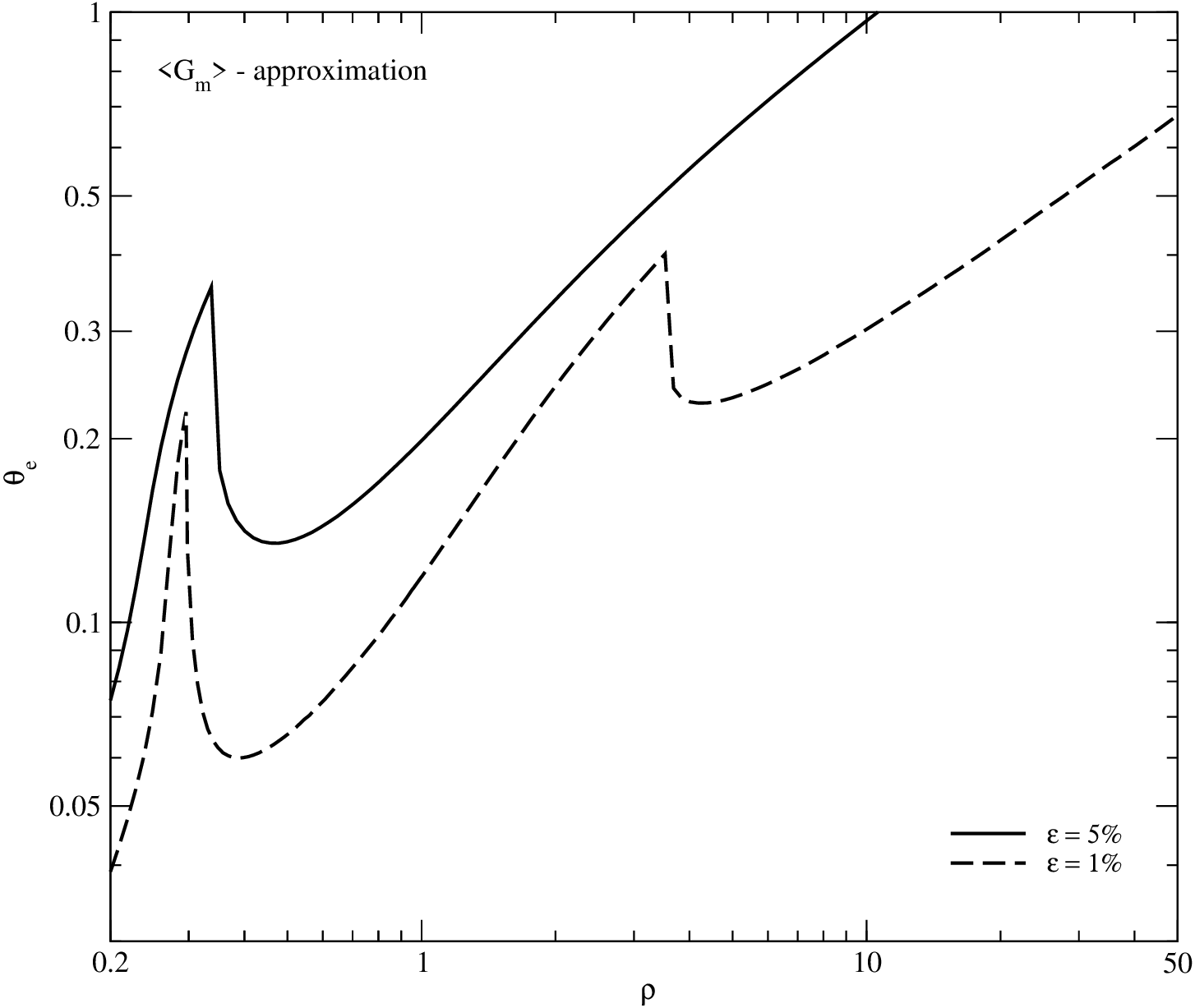}
\caption
{Range of applicability for Wien spectra and different $\rho=\Te/\Tg$: the
  electron temperature $\The$ is shown above which the relative difference between the
  considered analytic prediction and the numerical results becomes greater than
  $\epsilon$. The numerical results were compared with the approximation
  \eqref{eq:Gmono_inv}, the {\it direct} expansion \eqref{eq:Gfunctionres} up
  to fourth order in temperature and the {\it inverse} formula as given by
  Eq.~\eqref{eq:Ggeneralinv}.  }
\label{fig:Te_crit_Wien}
\end{figure}
\subsection{Range of applicability for the different approximations} 
\label{sec:app}
In the previous Sections we have derived different kinds of analytic
approximations for the DC emission coefficient. Each has advantages and
disadvantages. For example the approximation \eqref{eq:Gmono_inv} involves a
2-dimensional integral over the photon spectrum and the electron distribution,
which is numerically more expensive than the 1-dimensional integrals for the
direct expansion \eqref{eq:Gfunctionres} or the inverse approximation
\eqref{eq:Ggeneralinv}.
One obvious advantage of the inverse formula \eqref{eq:Ggeneralinv} over the
direct expansion \eqref{eq:Gfunctionres} is that only the first order
corrections are needed to obtain an excellent approximation up to fairly high
temperatures. Especially for numerical applications this is important, since
higher order derivatives of the spectrum may lead to significant difficulties.
However, it is necessary to investigate the range of applicability of each of
these approximations in more detail.

We have studied the accuracy of these approximations for Wien spectra more
extensively.
For this purpose we have compared the different approximations with the
results of the full numerical computation and, for a given $\rho=\Te/\Tg$,
determined the lowest values of the electron temperature $\The$ at which the
approximation deviates by more than $\epsilon$ from the numerical results.  
In Fig.~\ref{fig:Te_crit_Wien} we illustrate the outcome of these calculations
for the approximation \eqref{eq:Gmono_inv} based on the results for the
monochromatic case, the {\it direct} expansion \eqref{eq:Gfunctionres} up to
fourth order in temperature and the {\it inverse} formula from equation
\eqref{eq:Ggeneralinv}. 
The peaks in the curves for the direct expansion \eqref{eq:Gfunctionres} and
approximation \eqref{eq:Gmono_inv} appear where the relative difference to the
results of the full computation in each case changes sign.
For $\rho\lesssim 1$ the direct expansion breaks down very fast, while for
$\rho\gtrsim 1$ it works very well. The inverse formula works best for cases
close to $\rho=1$. Whenever induced effects are negligible the approximation
based on the results obtained for the monochromatic case is excellent,
in particular for $\rho\gtrsim 1$.

\section{Implications for the thermalization of spectral distortions
  of the CMB in the early Universe}
\label{sec:thermalization}
In the high redshift Universe ($z\gtrsim \pot{2.9}{5}$) injection of energy
into the medium leads to a \change{residual} $\mu$-type distortion of the CMB
\change{today} \citep{Suny70, Illarionov75, Illarionov75b, Danese82,
Burigana91, Hu93}.
\change{In this case the photon occupation number can be expressed as
$n(\nu)=1/[e^{h\nu/\kB\Tg+\mu}-1]$, with the dimensionless chemical potential
$\mu$, which in general is a function of frequency $\nu$.}
At low frequencies ($h\nu\lesssim \text{few}\times 10^{-3}\kB\Tg$) the
production of soft photons by the DC process returns the photon distribution
to a Planck spectrum after a very short time, whereas at high frequencies
\change{($h\nu\gtrsim 0.1\kB\Tg$) very few photons are newly emitted and}
Compton scattering is only able to establish a Bose-Einstein spectrum with
constant chemical potential.
\change{Therefore the time evolution of the high frequency spectrum depends
critically on the rate at which soft photons are created by the DC process and
then up-scattered by electrons via the Compton process.
}
%

The \change{current} limits on deviations of the CMB spectrum from a pure
blackbody as obtained with {\sc Cobe} \citep{Fixsen96,Fixsen02} constrain the
chemical potential today to obey $|\mu|< \pot{9}{-5}$, but this small residual
distortion could have arisen from a huge energy injection at sufficiently high
redshifts ($z\gtrsim 10^6-10^7$), \change{where until today the resulting
  spectral distortions were washed out by the combined action of DC soft
  photon production and Comptonization}.

Due to Compton scattering at these early stages the temperature of
the electrons is always very close to the \change{Compton} equilibrium
temperature in the given radiation field. For a $\mu$-type distortion with
$\mu\ll 1$ the difference between the radiation and electron temperature is
small and hence for estimates of the DC emission coefficient one may assume
$\Te\sim \Tg$.
\change{For small values of $\mu$ it is also sufficient to assume that the
  initial photon distribution is {\it very close} to a pure blackbody. As
  explained in Sect. \ref{sec:induced} stimulated DC emission will only affect
  the effective DC correction factor at the percent level. With
  $n(x)=1/[e^x-1]$ and Eqn. \eqref{eq:Ggeneralinv} $g^{\rm soft}_{\rm
    dc}\approx 1/[1+14.6\,\Thg]$, which implies that at high redshifts the
  thermalization of CMB spectral distortions will be less efficient, since the
  DC photon production rate is reduced compared to the Lightman-Thorne
  formula.}
For a temperature of $\kB\Tg\sim 4\,\text{keV}$, higher order corrections
lower the DC emissivity by $\sim 10\%$. This temperature corresponds to a
redshift of roughly $z\sim 10^7$. Therefore one can expect that
quantum-electrodynamical effects lead to significant corrections of the DC
photon production rate for large energy injection at redshifts $z\gtrsim
\text{few}\times 10^6$.
In addition, higher order temperature corrections to the Compton process may
also lower the thermalization efficiency.
This then will make the constraints placed on energy injection due
to exponentially decaying relict particles with short lifetimes ($\tX\lesssim
\text{few} \times 10^6\,\text{s}$) tighter and may affect the results obtained
within the standard treatment \citep{Hu1993a} \change{at a level $\sim
10\%$}.
\change{However, a more detailed computation is beyond the scope of this
work.

Higher order corrections to the DC process are negligible for the theoretical
computations of $y$-type distortions \citep{Zeldovich1969, Suny70,
  Illarionov75, Illarionov75b, Danese82, Burigana91, Hu93}, since these only
arise after energy release in the low redshift ($z\lesssim 10^4$) Universe,
where not only the temperature has become very small, but also Bremsstrahlung
dominates over DC emission.  }

\section{Conclusion}
\label{sec:Conc}
Double Compton emission in an isotropic, mildly relativistic thermal plasma
was investigated within the soft photon limit. Simple and accurate analytic
expression for the low frequency DC emission coefficient have been derived,
which extend the Lightman-Thorne approximation up to mildly relativistic
temperatures of the medium and should be applicable in a broad range of
physical situations.
In particular, expressions for initially monochromatic photons and Wien spectra
were given and discussed in detail.

It has been shown that the DC emissivity strongly decreases with higher mean
energy of the initial photons leading to a suppression of the total number of
newly created photons compared to the Lightman-Thorne approach. On the
other hand increasing the temperature of the electrons leads to an enhancement
of the DC emissivity.
If the photons and the electrons have similar temperatures, which is the case
in most physical situations close to full thermodynamic equilibrium, then
formulae \eqref{eq:Ggeneralinv} \change{should be} applicable up to $\kB T\sim
100\,$keV with an accuracy of better than a few percent.  Since only first
order corrections are necessary for this \change{approximation}, it is
generally most suitable for numerical applications.

\change{We also discussed possible consequences for the theoretical
  computations of the thermalization of $\mu$-type CMB spectral distortions at
  high redshifts ($z\gtrsim \text{few}\times 10^6$) and found that the new
  approximations could change the results at the level of a few percent,
  possibly up to $\sim 10\%$. However, these corrections are currently beyond
  the reach of observational possibilities.  }

\acknowledgement{We would like to thank the anonymous referee for comments.}

\newcommand{\jetp}{Soviet Physics JETP}

\onecolumn

\begin{appendix}
\section{Expansion of the relativistic Maxwell Boltzmann distribution}
\label{app:MB}
The low temperature expansion of \eqref{eq:relMBD} leads to
\beq\label{eq:relMBDexp}
\begin{split}
&f(E) 
=\frac{\Ne\,e^{-\xi}}{(2\pi\,\me^6\,\theta_{\rm e})^{3/2}}\,\left[
1 - \theta_{\rm e}\cdot\left(\frac{15}{8}-\frac{1}{2}\xi^2\right)+\theta^2_{\rm
e}\cdot\left(\frac{345}{128}-\frac{15}{16}\xi^2-\frac{1}{2}\xi^3+\frac{1}{8}\xi^4\right)
\right.
\\[1.0mm]
&\;\;\; -\theta_{\rm e}^3\cdot\left( \frac{3285}{1024} - \frac{345}{256}\,{\xi
}^2 - \frac{15}{16}\,{\xi }^3 - \frac{25}{64}\,{\xi }^4 + \frac{1}{4}{\,\xi
}^5 - \frac{1}{48}\,{\xi }^6 \right)\\[1.0mm]
&\;\;\;\;\; 
\left.
+\theta^4_{\rm e}\cdot\left( \frac{95355}{32768} -
\frac{3285\,{\xi }^2}{2048} - \frac{345\,{\xi }^3}{256} - \frac{855\,{\xi
}^4}{1024} - \frac{13\,{\xi }^5}{32} + \frac{51\,{\xi }^6}{128} - \frac{{\xi
}^7}{16} + \frac{{\xi }^8}{384}\right)
\right]
\Abst{,}
\end{split}
\eeq
with $\xi=\eta^2/2\,\theta_{\rm e}$ and $\eta=p/\me$.

\section{Squared matrix element for double Compton}
\label{appendix:M2}
For the DC differential cross section one has to calculate the squared matrix
element $|\mathcal{M}|^2=e^6\,X$ describing the DC process.  This calculation
was first performed by \citet{Mandl52}. In \citet{Jau76} one can
find the expression for the squared matrix element of the double Compton
process (pp. 235):
\beq
\begin{split}
X
&=2\left( a\,b - c\right)\big[ \left( a + b \right)
\left( 2+ x \right) -(a\,b - c) -8\big] - 2\,x\left[ a^2 + b^2 \right] 
-2\,\left[a\,b + c\left( 1 - x \right) \right]\rho
\\ 
& -8\,c +\frac{4\,x}{A\,B}\,\left[ \left( A + B \right) \,\left( 1 + x
\right) + x^2\,\left( 1 - z \right) + 2\,z - \left( a\,A + b\,B \right)
\,\left( 2 + \frac{\left( 1 - x \right) \,z}{x} \right) \right]
,
\end{split}
\eeq
where the following abbreviations have been used
\bsub\label{eq:MatAbr}
\beal
a&=\frac{1}{\kappa_0}+\frac{1}{\kappa_1} + \frac{1}{\kappa_2}&
b&=\frac{1}{\kappa'_0} + \frac{1}{\kappa'_1} + \frac{1}{\kappa'_2}& 
c&=\frac{1}{\kappa_0\,\kappa'_0}+ \frac{1}{\kappa_1\,\kappa'_1} + \frac{1}{\kappa_2\,\kappa'_2} 
\end{align}
\beal
x&={\kappa_0} + {\kappa_1} + {\kappa_2}&
z&=\kappa_0\,\kappa'_0+ \kappa_1\,\kappa'_1 + \kappa_2\,\kappa'_2 
\end{align}
\beal
A&={{\kappa}_0}\,{{\kappa}_1}\,{{\kappa}_2}& 
B&=\kappa'_0\,\kappa'_1\,\kappa'_2& 
\rho&=
\frac{\kappa_0}{\kappa'_0}+\frac{\kappa'_0}{\kappa_0}
+\frac{\kappa_1}{\kappa'_1}+\frac{\kappa'_1}{\kappa_1} 
+\frac{\kappa_2}{\kappa'_2} + \frac{\kappa'_2}{\kappa_2}
\Abst{.}
\end{align}
\esub
For the definitions of $\kappa_i,\,\kappa'_i$ we used those of the original paper
from \citet{Mandl52}:
\bsub\label{eq:KAPPAs}
\beal
\me^2\,\kappa_0&= -P\cdot K_0&
\me^2\,\kappa_1&= +P\cdot K_1&
\me^2\,\kappa_2&= +P\cdot K_2&
\\
\me^2\,\kappa'_0&= +P'\cdot K_0&
\me^2\,\kappa'_1&= -P'\cdot K_1&
\me^2\,\kappa'_2&= -P'\cdot K_2
\Abst{,}&
\end{align}
\esub
with the standard signature of the Minkowski-metric $(+\,-\,-\,-)$.
Assuming that the outgoing photon $\gamma(K_2)$ is {\it soft} compared to
$\gamma(K_1)$ one can expand $X$ into orders of the frequency $\omega_2$ and
keep only the lowest order term, i.e. terms of the order
$\mathcal{O}(\omega_2^{-2})$.  Similar expansions for the limits $\omega_0\ll
1$ and $\beta\ll 1$ can be performed.  Since the results of these expansions
are extremely complex and not very useful, we have omit them here.

\section{Numerical solution of the Boltzmann integrals}
\label{app:MC}
To solve the Boltzmann integrals we implemented two different programs, one
based on the {\sc Nag} routine {\tt D01GBF}, which uses an adaptive
Monte-Carlo method to solve multidimensional integrals, the other using the
{\sc Vegas} routine of the {\sc Cuba} Library\footnote{Download of the {\sc
    Cuba} Library available at: http://www.feynarts.de/cuba/}
\citep{Hahn2004}. The latter turned out to be much more efficient as it
significantly benefitted from importance sampling.
The expense and performance of the calculation critically depend on the
required accuracy. For most of the calculations presented here we chose a
relative error of the order of $\epsilon\sim 10^{-3}$. For the integrations
over different initial photon distributions we typically used $\omega_{\rm
  min}=10^{-4}\,\Thg$ and $\omega_{\rm max}=25\,\Thg$.

\subsection*{Integration over the electron momenta}
In order to restrict the integration region over the electron momenta for low
electron temperature we determined the maximal Lorentz factor, $\gamma_{\rm
  max}$, such that the change in the normalization of the electron
distribution was less than a fraction of the required accuracy.  Instead of
$p$ we used the variable $\xi=\eta^2/2\,\theta_{\rm e}$ with $\eta=p/\me$.

For high electron temperature it turned out to be more efficient to use the
normalization of the electron distribution itself as a variable. For this we
defined
\beal
\label{eq:N_int_var}
N(x)=\me^3 \int\nolimits_1^x \gamma\sqrt{\gamma^2-1}\,f(\gamma\me)\id\gamma
\Abst{,}
\end{align}
which for $x\rightarrow\infty$ becomes $N\equiv \Ne$. In actual calculations
one has to invert this equation to find the function $x(N)$. This was done
numerically before the integrations were performed for a sufficiently large
number of points ($n\sim 512$) such that the function $x(N)$ during the
integrations could be accurately represented via spline interpolation. Here it
is important to bear in mind, that $N(x)$ is rather steep for $x\sim 0$ and
$x\sim 1$.
 
\subsection{Definition of $I_k$}
Inserting the direct expansion of $G_{\rm m}$ in low energies of the initial
photon and electron temperature {Eq. \eqref{eq:Gmono_soft_w0_T_app}} into
Eq. \eqref{eq:Gmono_inv}
and collecting the different orders in $\theta_{\rm e}$, one can define the
functions $I_i$ of equation \eqref{eq:Gfunctionres} as
\bsub\label{eq:GFunc}
\beal
\label{eq:GFunc0}
I_0 &= \mathcal{D}_4 \\[1.0mm]
\label{eq:GFunc1}
I_1 &= 6\,\mathcal{D}_4 -\frac{21}{5}\,\frac{\Thg}{\The}\,\mathcal{D}_{5} \\[1.0mm]
I_2 &= 15\,\mathcal{D}_4 
-\frac{441}{10}\,\frac{\Thg}{\The}\,\mathcal{D}_{5}
+\frac{357}{25}\,\left[\frac{\Thg}{\The}\right]^2\,\mathcal{D}_6 
\\[1.0mm]
I_3 &= \frac{45}{4}\,\mathcal{D}_4 -\frac{8379}{40}\,\frac{\Thg}{\The}\,\mathcal{D}_{5} 
+\frac{5712}{25}\,\left[\frac{\Thg}{\The}\right]^2\,\mathcal{D}_6 
- \frac{7618}{175}\,\left[\frac{\Thg}{\The}\right]^3\,\mathcal{D}_{7} 
\\[1.0mm]
I_4 &= -\frac{45}{4}\,\mathcal{D}_4 
-\frac{3969}{8}\,\frac{\Thg}{\The}\,\mathcal{D}_{5} 
+\frac{8568}{5}\,\left[\frac{\Thg}{\The}\right]^2\,\mathcal{D}_6 
- \frac{34281}{35}\,\left[\frac{\Thg}{\The}\right]^3\,\mathcal{D}_{7}
+\frac{21498}{175}\,\left[\frac{\Thg}{\The}\right]^4\,\mathcal{D}_8
\Abst{,}
\end{align}
\esub
where we introduce the integrals $\mathcal{D}_{i}$ as 
\bsub\label{eq:HGH}
\beal
\mathcal{D}_{i} &= \int x^i\, n\id x
\end{align}
\esub
over the initial photon distribution $n(\nu)$.

\end{appendix}

\end{document}

%% file: Befehle.tex
\newcommand{\id}{{\,\rm d}}

\newcommand{\beq}{\begin{equation}}   %

\newcommand{\eeq}{\end{equation}}   %

\newcommand{\beqa}{\begin{eqnarray}}   %

\newcommand{\eeqa}{\end{eqnarray}}   %

\newcommand{\beal}{\begin{align}}

\newcommand{\enal}{\end{align}}

\newcommand{\bspl}{\begin{split}}

\newcommand{\espl}{\end{split}}

\newcommand{\bsub}{\begin{subequations}}

\newcommand{\esub}{\end{subequations}}

\newcommand{\bmulti}{\begin{multline}}   %

\newcommand{\beqm}{\begin{mathletters}}   %

\newcommand{\eeqm}{\end{mathletters}}   %

\newcommand{\Abst}[1]{\,#1}

\newcommand{\kB}{k_{\rm B}}

\newcommand{\me}{m_{\rm e}}

\newcommand{\Ne}{n_{\rm e}}

\newcommand{\Te}{T_{\rm e}}

\newcommand{\Tg}{T_{\gamma}}

\newcommand{\The}{\theta_{\rm e}}

\newcommand{\Thg}{\theta_{\gamma}}

\newcommand{\sigT}{\sigma_{\rm T}}

\newcommand{\nW}{n_{\rm W}}

\newcommand{\vek} [1]{\mbox{\boldmath${#1}$\unboldmath}}

\newcommand{\pd}{\partial}

\newcommand{\pAb}[2]{\frac{\displaystyle\pd #1}{\displaystyle\pd #2}}

\newcommand{\pot}[2]{#1 \times 10^{#2}}



%% file: paper.bbl
\begin{thebibliography}{}
\bibitem[Abramovitz \& Stegun(1965)]{Abramovitz1965} 
Abramovitz, M., \& Stegun, I.A., 1965, Dover Publications, Inc. New York
\bibitem[Burigana, Danese \& de Zotti(1991)]{Burigana91}
Burigana, C. Danese, L., \& de Zotti, G., 1991, \aap, 246, 49-58
\bibitem[Danese \& de Zotti(1982)]{Danese82}
Danese, L., \& de Zotti, G., 1982, \aap, 107, 39-42
\bibitem[{Fixsen} et al.(1996)]{Fixsen96}
{Fixsen}, D.J.,{Cheng}, E.S., {Gales}, J.M., {Mather}, 
J.C., {Shafer}, R.A.,{Wright}, E.L., 1996, \apj, 473, 576-587
\bibitem[Fixsen \& Mather(2002)]{Fixsen02}
Fixsen, D.J., \& Mather, J.C., 2002, \apj, 581, 817-822
\bibitem[Gould(1984)]{Gould84}
Gould, R.J., 1984, \apj, 285, 275-278
\bibitem[Hahn(2004)]{Hahn2004}
Hahn, T., 2004, hep-ph/0404043, download of the {\sc Cuba} Library available
from:
\\ 
http://www.feynarts.de/cuba/
\bibitem[Hu \& Silk(1993a)]{Hu1993a} 
Hu, W., \& Silk, J., 1993a, \prl, 18, 2661-2664
\bibitem[Hu \& Silk(1993)]{Hu93}
Hu, W., \& Silk, J., 1993, \prd, 48, 485-502
\bibitem[Illarionov \& Sunyaev(1975a)]{Illarionov75}
Illarionov, A.F., \& Sunyaev, R.A., 1975a, Sov. Astron., 18, 413-419
\bibitem[Illarionov \& Sunyaev(1975b)]{Illarionov75b}
Illarionov, A.F., \& Sunyaev, R.A., 1975b, Sov. Astron., 18, 691-699
\bibitem[Jauch \& Rohrlich(1976)]{Jau76}
Jauch, J.M., \& Rohrlich, F., 1976, Springer-Verlag
\bibitem[Lightman(1981)]{Light81}
Lightman, A.P., 1981, \apj, 244, 392-405
\bibitem[Mandl \& Skyrme(1952)]{Mandl52}
Mandl, F., \& Skyrme, T.H.R., 1952, Proc. Roy. Soc., A215, 497
\bibitem[Pozdnyakov et al.(1983)]{pozdnyakov1983}
  Pozdnyakov, L.A., Sobol, I.M., Sunyaev R.A., 1983, Astrophys. \&
  Space Phys. Rev., ed. R.A. Sunyaev, Harwood Academic
  Publishers, Chur, vol. 2, 189
\bibitem[Svensson(1984)]{Svensson84}
Svensson, R., 1984, \mnras, 209, 175-208
\bibitem[Sunyaev \& Zeldovich(1970)]{Suny70}
Sunyaev, R.A., \& Zeldovich, Ya. B., 1970, \apss, 7, 20-30
\bibitem[Thorne(1981)]{Thorne81}
Thorne, K.S., 1981, \mnras, 194, 439-473
\bibitem[Zeldovich \& Sunyaev(1969)]{Zeldovich1969} Zeldovich, Y.~B., 
\& Sunyaev, R.~A.\ 1969, \apss, 4, 301 
\end{thebibliography}
